\documentclass[onecolumn,superscriptaddress,amsmath,amssymb,aip]{revtex4-1}
\usepackage{textcomp}
\usepackage{wasysym}
\usepackage{physics}
\usepackage{blindtext}
\usepackage{graphicx}  
\usepackage{dcolumn}  
\usepackage{bm}
 
\usepackage{xcolor}
\color{black}
\usepackage[colorlinks=blue]{hyperref}
\hypersetup{linkcolor=blue,anchorcolor=black,citecolor=blue,filecolor=cyan, menucolor=red,runcolor=cyan,urlcolor=magenta}
\usepackage{cleveref}

\begin{document}
\title{Magnetic-field-induced incommensurate to collinear spin order transition in NiBr$_{2}$}

\author{S. Babu}
\affiliation{%
School of Materials Science and Technology, Indian Institute of Technology, Banaras Hindu University, Varanasi-221005, India}%

\author{K. Proke\v{s}}
\email{prokes@helmholtz-berlin.de}
\affiliation{%
Helmholtz-Zentrum f\"{u}r Materialien und Energie, Hahn-Meitner-Platz 1, D-14109, Germany
}%
\author{Y. K. Huang}
\affiliation{%
Van der Waals-Zeeman Institute, University of Amsterdam, Science Park 904, The Netherlands,
}%
\author{F. Radu}
\affiliation{%
Helmholtz-Zentrum f\"{u}r Materialien und Energie, Albert-Einstein-Stra\ss{}e 15, D-12489, Germany
}%
\author{S. K. Mishra}
\email{shrawan.mst@iitbhu.ac.in}
\affiliation{%
School of Materials Science and Technology, Indian Institute of Technology, Banaras Hindu University, Varanasi-221005, India}% 

\date{\today}

\begin{abstract}
The triangular spin lattice of NiBr$_{2}$ is a canonical example of a frustrated helimagnet that shows a temperature-driven phase transition from a collinear commensurate antiferromagnetic structure to an incommensurate spin helix on cooling. Employing neutron diffraction, bulk magnetization, and magnetic susceptibility measurements, we have studied the f\hspace*{.5pt}ield-induced magnetic states of the NiBr$_{2}$ single crystal. Experimental f\hspace*{.5pt}indings enable us to recapitalize the driving forces of the spin spiral ordering in the triangular spin-lattice systems, in general. Neutron diffraction data conf\hspace*{.5pt}irms, at low temperature below T$_{{\rm m}}$ = 22.8(1) K, the presence of diffraction satellites characteristic of an incommensurate magnetic state, which are symmetrically arranged around main magnetic reflections that evolve just below T$_{{\rm N}}$ = 44.0(1) K. Interestingly, a f\hspace*{.5pt}ield-induced transition from the incommensurate to commensurate spin phase has been demonstrated that enforces spin helix to restore the high temperature compensated antiferromagnetic structure. This spin reorientation can be described as a spin-flop transition in the (\hbox{$a$--$b$}) basal plane of a triangular spin lattice system.  These f\hspace*{.5pt}indings offer a new pathway to control the spin helix in incommensurate phases that are currently considered having high technical implications in the next-generation data storage devices.

\end{abstract}
\maketitle

\section{\label{sec:level1}Introduction}

Noncollinear spins are generic to the systems with spin spirals, spin helicoids, canted spins, and ferromagnetic (antiferromagnetic) domain walls. Such spin systems have attracted signif\hspace*{.5pt}icant interest in recent research activities since they are potential candidates for ultra-dense magnetic storage devices \textcolor{blue}{\cite{Fert152,Sampaio839,Beach611}}. Noncollinear spin helix mainly evolve in the presence of inversion symmetry breaking in such systems and yields various unusual physical phenomena like electric excitation of magnon \textcolor{blue}{\cite{Ramirez406,Tokura510}} and magnetic skyrmions \textcolor{blue}{\cite{Muhlbauer,Rossler}}. Various microscopic driving mechanisms would produce noncollinear spin ordering, and among them, Dzyaloshinskii--Moriya interaction (DMI) is the most common in many of these systems \textcolor{blue}{\cite{Dzyaloshinskii658,Moriya760}}.

Noncollinear spin arrangements in helimagnets also result from the presence of competing ferromagnetic and antiferromagnetic Heisenberg interactions between spins \textcolor{blue}{\cite{Mostovoy817,Masuda,Zinke,Villain977}}. Considering exchange interactions model in a magnetic f\hspace*{.5pt}ield of intensity H, the Hamiltonian \textbf{H} for a triangular lattice can be expressed through the following Eq. (1)
\begin{equation}
\textbf{H}=-J_{1}\Sigma_{<i,j>}S_{i}.S_{j}-J_{2,3}\Sigma_{<<i,j>>}S_{i}.S_{j}-H\Sigma_{i}S_{i}
\end{equation}

where, $\Sigma_{<i,j>}$ and $\Sigma_{<<i,j>>}$represent the sum over the nearest neighbor (\textit{nn}) and third nearest neighbor (\textit{nnn}), respectively \textcolor{blue}{\cite{Mostovoy817}}. For  ferromagnetic state ($J_{1}>0$) with only \textit{nn} interaction, the ground state of such system would be commensurate to the underlying lattice. However, on the other hand, for competing interactions such as $J_{1}>0$ and the AF third-neighbor interaction $J_{3}<0$, an incommensurate ground state is more favorable \textcolor{blue}{\cite{Mostovoy817,Reatto979}}. In other words, when \textit{nnn} interaction become dominant, the ground state is often an incommensurate spiral structure \textcolor{blue}{\cite{Mostovoy817,Reatto979,Kimura1007,Fishman1109,Fishman_prb}}. In such systems, the magnitude of \textit{nnn} interaction is larger than the \textit{nn} interaction. Although the ordered magnetic structures are normally robust, exciting phase diagrams as a function of temperature and applied magnetic f\hspace*{.5pt}ield are often encountered \textcolor{blue}{\cite{Mostovoy817,Reatto979,Kimura1007,Fishman1109,Fishman_prb}}.

Transition metal dihalides have long been studied as prototypes of antiferromagnets with a triangular lattice \textcolor{blue}{\cite{Ward1282,Tokura1311,Tokura1411}}. Most of these investigations were focused on the coupling between magnetism and ferroelectricity in MnI$_{2}$ and NiBr$_{2}$ \textcolor{blue}{\cite{Tokura1311,Tokura1411}}. The latter system has been considered as a model system for spin triangular lattice \textcolor{blue}{\cite{Robbins1588,Meriel1680,Terrier1982}}. When the temperature reduces, the NiBr$_{2}$ displays a paramagnetic-antiferromagnetic phase transition at $T_{{\rm N}}$ = 44.0(1) K and then antiferromagnetic-incommensurate spiral transition at $T_{{\rm m}}$ = 22.8(1) K \textcolor{blue}{\cite{Robbins1588,Meriel1680,Terrier1982}}.

The ground state of the incommensurate spiral exhibits a threefold degeneracy with respect to the three equivalent wave vectors of the lattice. This threefold degeneracy is predicted to be  a source of exotic ordered states, namely, multiple-\textit{q} state \textcolor{blue}{\cite{Okubo2012}}.  This multiple-\textit{q} state may not be favorable  at lowest temperature, but they may be stabilized at moderate temperature by thermal fluctuations and finite external magnetic fields. Such multiple-\textit{q} state corresponds to the Skyrmion lattice state which may occur in such a system, as considered theoretically by Okubo et al. Phys. Rev. Lett. \textbf{108}, 017206 (2012) \textcolor{blue}{\cite{Okubo2012}}.

In this work, we address the formation of a multiple-\textit{q} state in the NiBr$_{2}$ triangular frustrated Heisenberg system.  We have investigated various properties of the incommensurate spiral phase in the single crystal NiBr$_{2}$ using dc(ac) magnetization and magnetic susceptibilities, neutron diffraction, and small angle neutron scattering. The experimental f\hspace*{.5pt}indings suggest that NiBr$_{2}$ has a helimagnetic ground state in good agreement with the previous observations \textcolor{blue}{\cite{Tokura1311,Tokura1411,Robbins1588,Meriel1680}}. The ground state is reached upon reducing the temperature through a commensurate phase. In an applied magnetic field and at finite temperature, we could not find evidence for multiple-\textit{q} state occurrence. Instead, we found that the system responds to the structural symmetry of 120$^\circ$ by populating the three equivalent wave vectors in the incommensurate state. Moreover, we emphasize on a new transition from the ground state incommensurate spiral phase to commensurate one under magnetic f\hspace*{.5pt}ield (H$>$ 4 T) applied perpendicular to the $c$-axis. The f\hspace*{.5pt}ield-induced phase appears to be the same as the high temperature phase, i.e., NiBr$_{2}$ exhibits a spin order transition, where helical spins can regain high temperature antiferromagnetic phase even the system temperature is at well below $T_{\mathrm{m}}$. The unique response of spin-helix phase under external f\hspace*{.5pt}ield (H) indicates that even a slight difference in the magnetic interactions is critically reflected in the respective response in frustrated magnets.

\section{\label{sec:level2}METHODS AND EXPERIMENTAL DETAILS}

High-quality single crystals of NiBr$_{2}$ were grown at the University of Amsterdam, using a self-flux growth method starting from a stoichiometric mixture of Ni, and Br, as reported in Tokunaga et al. and Kurumaji et al. \textcolor{blue}{\cite{Tokura1311,Tokura1411}}. The samples have been studied with the average weight of 10 mg. The single crystals had typical dimensions of approximately $4 \times 4 \times 2\, \mathrm{mm}^{3}$. NiBr$_{2}$ crystals are sensitive to air/moisture; therefore, they are kept in the vacuum to prevent any degradation of the sample's quality. NiBr$_{2}$ crystal forms naturally layered shape with the \textit{c}-axis perpendicular to planes that eff\hspace*{.5pt}iciently cleave. The nickel ions occupy a single site in the Bravais lattice as shown in Fig.~\href{Color online}{\textcolor{blue}{1(a)}}. Magnetic measurements at various temperatures in the range 2~K to 300~K were carried out using the MPMS 7T device and PPMS 14T (Quantum Design) systems, in f\hspace*{.5pt}ields up to 14 T, applied within and out of the plane directions. The measurements and analyses of the dc(ac) susceptibility have been conducted on samples that were f\hspace*{.5pt}ixed to quartz sample holders using a diamagnetic glue.

NiBr$_{2}$ crystal originating from the same batch used for both magnetic measurements as well as for neutron diffraction experiments, which were carried out on the diffractometers E2 and E4 and the small-angle neutron scattering instrument V4 at the BER II reactor of the Helmholtz-Zentrum Berlin (HZB) \textcolor{blue}{\cite{Yokoachiya1717}}. In the part, magnetic f\hspace*{.5pt}ields up to 5 T were applied along and perpendicular to the trigonal axis of the single-crystalline NiBr$_{2}$ using a horizontal-f\hspace*{.5pt}ield cryomagnet that restricts signif\hspace*{.5pt}icantly the scattering geometry. In particular, this makes impossible to survey a large portion of the reciprocal space due to dark angles imposed by the construction of the magnet coils. E2 and E4 instruments use a pyrolytic graphite (002) monochromator selecting the neutron wavelength $\lambda = 2.4$ \AA. The data were collected with two-dimensional position sensitive $^{3}$He detectors that were of different sizes at respective instruments. The f\hspace*{.5pt}ield- and temperature-dependent scans were performed by measuring the peak intensity while ramping the f\hspace*{.5pt}ield/temperature. The integrated peak intensity was found by summing the counts in a small area of the detector surrounding the reflection. On E2 and E4, in order to ref\hspace*{.5pt}ine the magnetic structure of NiBr$_{2}$, we have collected data mainly in the (\textit{hhl}) plane. However, a signif\hspace*{.5pt}icant portion below and above the scattering plane has been covered as well. In addition, on E4, we have collected data also in the (\textit{hk0}) plane. 

The crystal and magnetic structure ref\hspace*{.5pt}inements were performed with the program Fullprof \textcolor{blue}{\cite{Roisnel1801,Sears1992}}. The nuclear scattering lengths b(Ni) = 1.03 fm and b(Br) = 0.6795 fm were considered. For the absorption correction (Gaussian integration), we used the absorption coeff\hspace*{.5pt}icient $\mu = 0.12\, \mathrm{cm}^{-1}$. No secondary extinction correction has been applied. $\frac{\lambda}{2}$ f\hspace*{.5pt}ilters were used at both E2 and E4 instruments imply residual higher-order wavelength contamination at a level of less than $10^{-4}$. 

In addition, the small-angle neutron scattering measurements were carried out with instrument V4 at HZB that possess a large position-sensitive detector movable at the distance between 1 and 16 m from the sample. The V4, where we have also applied the magnetic f\hspace*{.5pt}ield along the basal plane of the crystal, has been used in a conf\hspace*{.5pt}iguration that covered $0.04\, \mathrm{nm}^{-1} < Q < 1.3\, \mathrm{nm}^{-1}$.

\section{\label{sec:level3}RESULTS AND DISCUSSIONS}

\subsection{\label{sec:results}Magnetization and Susceptibility}

Figure \href{Color online}{\textcolor{blue}{2(a)}} illustrates the temperature dependence of the magnetic susceptibility $\chi= M/H$ of NiBr$_{2}$, where, the $M$ is magnetization, and the $H$ is magnetic f\hspace*{.5pt}ield ranging between 1.0 and 14.0 T, which was applied perpendicular to the \textit{c}-axis. In the case of NiBr$_{2}$ at high temperatures where $\chi= M/H$ measured the magnetization is directly proportional to the applied magnetic field (i.e. linear in the field), the evaluated M/H values are identical to the magnetic susceptibility at different fields are insensitive to the magnetic field strengths.

It is evident, that $\chi(T)$ is f\hspace*{.5pt}ield independent for temperatures above \AC80~K. At lower temperatures, two well-distinguished anomalies can be discerned marking the magnetic phase transition temperature that were determined from maxima of $\frac{\partial T.\chi(T)}{\partial T}$. 

At 1.0 T magnetic f\hspace*{.5pt}ield, the transition between the paramagnetic and an antiferromagnetic state is detected at $T_{{\rm N}}$ = 44.6(1)~K. A closer inspection reveals that for increasing the magnetic f\hspace*{.5pt}ield, this transition shifts toward lower temperatures. The second transition between AF and helical spin order takes place at $T_{{\rm m}}$ = 21.8 (1) K. As it is evident from Fig.~\href{Color online}{\textcolor{blue}{2(a)}}, this anomaly also shifts toward lower temperatures as a function of the magnetic f\hspace*{.5pt}ield. 

The temperature dependence of $1/\chi(T)$ and their f\hspace*{.5pt}its to a modif\hspace*{.5pt}ied Curie--Weiss law are shown in inset panels of Fig.~\href{Color online}{\textcolor{blue}{2(a and b)}}. The modif\hspace*{.5pt}ied Curie--Weiss law can be expressed by Eq.(2):
\begin{equation}
\chi=C/(T-\theta_{\mathrm{p}}) + \chi_{0}
\end{equation}
where, C represents the Curie constant and $\theta_{\mathrm{p}}$ the paramagnetic Curie temperature. $\chi_{0}$ denotes the temperature-independent terms. The excellent quality of f\hspace*{.5pt}its conf\hspace*{.5pt}irms the modif\hspace*{.5pt}ied Curie--Weiss behavior for the system above $T > 70$ K. The best f\hspace*{.5pt}it of data acquired with 1.0 T ($T> 150$ K) yields the effective moment of $\mu_{{\rm eff}}$ = 2.76 (1) $\mu_{\mathrm{B}}$. This value is approximately the same as the effective moment expected for the S = 1, 3d$^8$ electronic conf\hspace*{.5pt}iguration of Ni ions $(\sqrt{8} = 2.83)$. The effective moment determined from data taken at 14.0 T amounts to 2.66 (1)$\mu_{\mathrm{B}}$. The fitted values of $\theta_{\mathrm{p}}$ amount to 22.2(1) K and 28.1 (1) K for the data recorded 1.0 T and 14.0 T, respectively.

Figure \href{Color online}{\textcolor{blue}{2(b)}} shows the temperature dependence of magnetic susceptibility measured for the various f\hspace*{.5pt}ield applied along the \textit{c}-axis. Also for this orientation, the magnetic susceptibility at higher temperatures is f\hspace*{.5pt}ield independent with anomalies visible only at the lower temperatures. The effective magnetic moment and the paramagnetic Curie temperature are very similar to the other orientation ($\mu_{\mathrm{eff}}$ =2.74 (1)$\mu_{\mathrm{B}}$ and $\theta_{\mathrm{p}}$ = 21.3 (1) K, respectively). The best fit to data taken at 14 T applied along the c-axis is shown in the inset of Fig. \href{Color online}{\textcolor{blue}{2(b)}}.

For $T > 70$ K, the similarity between magnetic susceptibilities with f\hspace*{.5pt}ield applied along and perpendicular to the \textit{c}-axis suggests that the anisotropy in paramagnetic state of NiBr$_{2}$ is negligible. However, this does not hold true for the state below $T_{{\rm N}}$ as magnetization (\hbox{M--H}) curves taken for f\hspace*{.5pt}ield applied perpendicular and along the \textit{c}-axis had shown entirely different magnetization behaviors, as depicted in Fig.~\href{Color online}{\textcolor{blue}{3(a and b)}}. Selectively, for the case when applied magnetic f\hspace*{.5pt}ield within the (\hbox{\hbox{$a$--$b$}}) basal plane, a clear f\hspace*{.5pt}ield-induced transition that shifts with increasing temperature to lower f\hspace*{.5pt}ields is evident. At 3~K, the magnetization step associated with the transition amounts to about 0.036 $\mu_{\mathrm{B}}$/Ni. This is in good agreement with the magnetic susceptibility data. For the second geometry, when f\hspace*{.5pt}ield was applied along the \textit{c}-axis, no such trends were observed.

Now we discuss the insights on phase transitions through measurements of the temperature and magnetic f\hspace*{.5pt}ield dependence of ac magnetic susceptibility, which is a complex value and expressed as $\chi_{ac}$ = $\chi'-i\chi^{\prime\prime}$. The real component, $\chi'$, is related to the reversible magnetization process and always in-phase with the oscillating f\hspace*{.5pt}ield. However, imaginary component, $\chi^{\prime\prime}$, is related to losses during the irreversible magnetization process. 

The measurements of ac susceptibilities were carried out for multiple values of applied f\hspace*{.5pt}ields as a function of temperature (measured with H = 1 mT and f = 99 Hz). To minimize the demagnetization effects, the dc magnetic f\hspace*{.5pt}ields were applied within the basal plane (\hbox{\hbox{$a$--$b$}}). Fig.~\href{Color online}{\textcolor{blue}{4(a)}} illustrates the temperature dependence of the ac susceptibilities measured in the vicinity of both transitions. Experimental results over the f\hspace*{.5pt}ield range from 0.1 to 2.0 T show the emergence of maxima across both phase transitions. Over the $T_{{\rm m}}$, ac susceptibility displays a maxima with respect to the temperature; the position of which trending towards higher side (while the amplitude of ac susceptibility decreases) with reducing the strength of applied magnetic f\hspace*{.5pt}ields, similar as in Ref. \textcolor{blue}{\cite{Tokura1411}}. The variation in peak heights and temperatures for various dc magnetic f\hspace*{.5pt}ields shows a monotonic variation with strength of the applied dc f\hspace*{.5pt}ields.

Figure \href{Color online}{\textcolor{blue}{4(b)}} shows the ac susceptibilities as a function of applied magnetic f\hspace*{.5pt}ields for various constant temperatures that range from 5~K to 80~K. At $T=80$~K, the ac susceptibility exhibits a line-shaped curve with low amplitude. At $T = 25$~K; a peak can be seen close to the upper phase boundary of the antiferromagnet state as also seen in Fig. \href{Color online}{\textcolor{blue}{2(a)}}. At even lower temperature near T=20~K, an increase in ac susceptibility peak amplitude is clearly noticed for applied low f\hspace*{.5pt}ields 0.5 T $<$ H $<$ 2.5 T. This modif\hspace*{.5pt}ied ac susceptibility magnitude is also accompanied by a systematic \hbox{enhancement} in the peak f\hspace*{.5pt}ield, which would correspond to the boundary of incommensurate phase in this region.

Reapeating the f\hspace*{.5pt}ield dependence of ac susceptibilities at various temepratures across incommensurate phase result a $\chi$(H,T) curve as illustarted in Figure \href{Color online}{\textcolor{blue}{4(b)}}. Below $T_{{\rm m}}$, a hysteresis nature of $\chi$(H,T) curves clearly evident, which acquired for different temepratures, conf\hspace*{.5pt}irming the helix nature of noncollinear nature of spins. 

The shift in peak positions as a function magnetic f\hspace*{.5pt}ield represents the torque required to reverse the helical spin upon flipping the polarization of applied magnetic f\hspace*{.5pt}ield. In this way, it might be possible to
identify the spin reorientation processes, which are responsible for
the unusual field dependence in the ac susceptibility and subsequently investigate the magnetic field effects on the physical interactions driving these phase transition processes. The results reported above indicate the emergence of field-induced incommensurate to commensurate phase transition (as demonstrated further below by neutron scattering),
a generic response that occurs only acros the low temperature boundary of these phases. Note that in Ref. \textcolor{blue}{\cite{Tokura1411}}, this is ascribed to a transition from cycloidal to collinear AF order. The ac susceptibility peak positions in (H,T) dimensions over
the typical ranges map directly onto the field-induced
modulation in incommensurate phase. It should be noted
that at zero field, ac susceptibility diminishes monotonically over the range of temperatures between 5 K
and 20 K.

Figure \href{Color online}{\textcolor{blue}{4(c)}} shows the comparison between $\chi_{{\rm dc}}$ and  $\chi'$$_{{\rm ac}}$, both measured at $H_{{\rm dc}}$ = 2 kOe. Their respective temperature derivatives are depicted in the inset of Fig.~\href{Color online}{\textcolor{blue}{4(c)}}. The $\chi'$$_{{\rm ac}}$ provides a precise determination of the long-range magnetic phase ordering temperatures, which can be determined from the $\chi'$$_{{\rm ac}}$ or $d\chi'$$_{{\rm ac}}$$/dT$ maxima. The f\hspace*{.5pt}ield and temperature dependence of the ac susceptibility at 20~K is more complicated. In this \hbox{region}, the ac susceptibility peak amplitudes f\hspace*{.5pt}irst increase before dropping again with dc f\hspace*{.5pt}ields. In contrast, the susceptibility peak temperatures increase monotonically with f\hspace*{.5pt}ield, tracking closely along the phase boundaries across $T_{{\rm m}}$ regimes determined from the experimental values of \vspace*{1.5pt} $\frac{\partial M(T)}{\partial T}$ and $\frac{\partial M(H)}{\partial H}$. This behavior of $\chi_{{\rm ac,dc}} (H,T)$ is a def\hspace*{.5pt}ining characteristic of such transitions, specif\hspace*{.5pt}ically reflecting the combination of ferromagnetic and antiferromagnetic interactions in the triangular NiBr$_{2}$ system, which leads to an intricated influence of applied f\hspace*{.5pt}ields as a function of the changing spin conf\hspace*{.5pt}igurations.

\subsection{\label{sec:N Diffraction} Neutron diffraction on NiBr$_{2}$ crystal}

NiBr$_{2}$ crystallizes in the CdCl$_{2}$ structure, which has the rhombohedral space group \textit{D$^{5}_{3d}$} \textcolor{blue}{\cite{Ward1282,Tokura1311,Tokura1411,Robbins1588,Meriel1680}}. In this space group, the Ni and Br atoms are situated at the Wyckoff positions 3\textit{a}(0,0,0) and 6\textit{c}(0,0,\textit{z}) with 3~\textit{m} and $-$3~\textit{m} local symmetries, respectively \textcolor{blue}{\cite{Robbins1588,Meriel1680}}. The lattice constants are a = 3.723 \AA \hspace*{.5pt}  and c = 18.340 \AA. 
The space group leads to specif\hspace*{.5pt}ic extinction conditions, for instance, the reflections \textit{hkil}: \textit{$-$h$+$k$+$l} = 3\textit{n} and the reflections \textit{hki0}: \textit{$-$h$+$k} = 3\textit{n}. Above $T_{{\rm N}}$, in the course of neutron diffraction experiments, we could observe only nuclear Bragg reflections.

Figure \href{Color online}{\textcolor{blue}{5(a)}} illustrates the (\textit{hhl}) diffraction pattern recorded at 50~K. As it is evident that only (110), (003), (006) and (009) Bragg reflections are observed (along with signal originating from the aluminum sample holder/cryostat)\textcolor{blue}{\cite{Robbins1588,Meriel1680}}. There were no unexpected Bragg reflections observed. The ref\hspace*{.5pt}inement of observed Bragg reflections leads to a reasonable agreement with literature data with the only structural free parameter z$_{Br}$ = 0.268. As the temperature is decreased below $T_{{\rm N}}$ (but above $T_{{\rm m}}$), new Bragg reflections associated with AF order are observed \textcolor{blue}{\cite{Robbins1588,Meriel1680}}. These can be indexed with a propagation vector $q_{{\rm c}}$ = (0 0 3/2) as shown in Fig.~\href{Color online}{\textcolor{blue}{5(b)}} and which is in well agreement with the literature \textcolor{blue}{\cite{Robbins1588,Meriel1680}}. The diffraction pattern recorded at 26~K conf\hspace*{.5pt}irms that (0 0 3/2), (0 0 9/2), and (0 0 15/2) magnetic reflections show non-zero intensity. 
These magnetic satellite structures around the (003) m, (006) m, and (009) m corroborate the antiferromagnetic ordering of Ni spins in (\hbox{\hbox{$a$--$b$}}) basal plane \textcolor{blue}{\cite{Robbins1588,Meriel1680,Yokoachiya1717, Stock}}.

As the magnitude of incommensurate component at lower temperature is expected to be very small, special care has been paid to improve the resolution of the diffractometer. We have utilized high wavelength neutrons with $\lambda=2.4$ \AA  produced by a pyrolytic graphite monochromator and graphite filters to remove the  $\lambda/2$ contamination.  A reduced focus lead on one side to lower intensities but also to decreased vertical divergence needed to separate the magnetic reflections in the vertical direction. To improve the signal to noise ratio we have closed both the horizontal and vertical slits as much as possible while still irradiating the sample completely \textcolor{blue}{\cite{Yokoachiya1717,Stock}}. These scans show that the component of propagation vector along the \textit{c}-axis remains commensurate with a value k$_{{\rm z}}$ = 3/2. However an incommensurate component exists in the (\hbox{\hbox{$a$--$b$}}) basal plane, which is oriented along $<110>$ direction. From 4.2~K to 21.8~K the propagation vector has been found to be temperature dependent. The k$_{{\rm z}}$ component remains commensurate to a value 3/2. The in-plane component is always along $<110>$ direction. The absence of any second, third, or higher harmonics (Fig.~\href{Color online}{\textcolor{blue}{5(c)}}) proves that NiBr$_{2}$ orders with a helical structure within the (\hbox{\hbox{$a$--$b$}}) basal plane in zero f\hspace*{.5pt}ield at $T$ = 4.2~K. The planes remain coupled antiferromagnetically along the c-axis direction \textcolor{blue}{\cite{Robbins1588,Meriel1680,Stock}}. The magnetic satellites regain its shape similar to commensurate antiferromagnetic phase upon applying an in-plane magnetic f\hspace*{.5pt}ield (H $\leq$ 4~T) as shown in Fig.~\href{Color online}{\textcolor{blue}{5(d)}}. This spin order transition of magnetic satellite peak is reported in the incommensurate phase for a triangular spin systems. A signif\hspace*{.5pt}icant change in the diffraction patterns reported at T=21.1K, as the $q_{{\rm c}}$ Bragg reflections split (see Fig.~\href{Color online}{\textcolor{blue}{5(c)}}). The estimated propagation is of the $q_{{\rm i}}$ = (q$_{{\rm h}}$ q$_{{\rm h}}$ 3/2), with q$_{{\rm h}}$ $\approx$ 0.03. There are no other magnetic reflections that described by q$_{2}$ are visible at 2~K, zero f\hspace*{.5pt}ield. These scans show that the component of the propagation vector along the $c$-axis remains commensurate with a value k$_{{\rm z}}$ value 3/2 \textcolor{blue}{\cite{Robbins1588,Meriel1680,Yokoachiya1717,Stock}}.

Magnetic measurements have indicated for the f\hspace*{.5pt}ield applied within the (\hbox{\hbox{$a$--$b$}}) plane, a magnetic phase transition around 3 T. Figure \href{Color online}{\textcolor{blue}{5(d)}} shows the pattern recorded at 2~K and 4 T. As can be seen, it is identical to the pattern recorded at 26~K (see Fig.~\href{Color online}{\textcolor{blue}{5(b)}}). This can be at best seen from the insets of the respective panels. In the case of data taken at 50~K, one can also notice a detectable short-range ordered signal. Such f\hspace*{.5pt}ield-induced spin reorientation in single crystal of incommensurate NiBr$_{2}$ is observed with a magnetic f\hspace*{.5pt}ield applied perpendicular to \textit{c}-axis. Spin reorientation occurred continuously with increasing f\hspace*{.5pt}ield. At sample temperatures of 4.2~K and 22~K, a complete spin reorientations were observed with critical f\hspace*{.5pt}ields (H $\geq$ 4~T). These f\hspace*{.5pt}indings will be further examined within the framework of spin-flop coupling in the next section.

Further, we have recorded the quality of crystalline lattice, rocking curves of NiBr$_{2}$ crystal, concentrating on a small region of reciprocal space at a constant \textit{q}-value. Figure \href{Color online}{\textcolor{blue}{6(a)}} shows the rocking curve measured through the (0 0 3) nuclear Bragg reflection at 55~K in the \hbox{absence} of any applied magnetic f\hspace*{.5pt}ields. The full width of the reflection at half maximum (FWHM) amounts to 1.9\textdegree, which is about four times more than the resolution of the used instrument. A comparative rocking curve through the magnetic (0 0 3/2) Bragg reflection were recorded at 23~K in zero f\hspace*{.5pt}ield, as depicted in Fig.~\href{Color online}{\textcolor{blue}{6(b)}}. The FWHM for (0 0 3/2) reflection is larger than the corresponding nuclear reflection (0 0 3). The shape is similar to the nuclear reflection suggesting that magnetic ordering takes place in the whole volume of the sample, without any influence of crystal mosaicity.

Figure \href{Color online}{\textcolor{blue}{6(c)}} shows the temperature dependence of the magnetic (0 0 3/2) Bragg reflection recorded with zero f\hspace*{.5pt}ield upon cooling at the top of the reflection. The solid red line represents the best f\hspace*{.5pt}it of the experimental data using Eq. (3):
\begin{equation}
I(T)=b+I_{0}(1-\frac{T}{T_{{\rm N}}})^{2\beta}
\end{equation} 
where, b denotes the background intensity, $I_{0}$ represents the intensity at 0~K, and $\beta$ is the critical parameter related to dimensionality of the system. The best f\hspace*{.5pt}it to this empirical formula, which is valid in the critical region near the magnetic phase transition, leads to a good description of the data above 40~K. The magnetic phase transition occurs at $T_{{\rm N}}= 44.5$ (6) K with $\beta$ = 0.30 (1). However, the observation of scattered intensity (see the non-negligible intensity above the background level in the Fig.~\href{Color online}{\textcolor{blue}{6(c)}} highlighted by the dashed line) above the $T_{{\rm N}}$ points to the presence of critical scattering in this material. The second-order transition at $T_{{\rm m}}$ takes place at 21.6 (5) K. Below this temperature, the intensity is signif\hspace*{.5pt}icantly above its high temperature background level. However, in this case, as it is shown in Fig.~\href{Color online}{\textcolor{blue}{6(c)}}, the intensity originates from new Bragg reflections that contribute to the original position of the (0 0 3/2) Bragg reflection due to a f\hspace*{.5pt}inite resolution of the E4 instrument and spread of the intensity originating from different magnetic Bragg reflections.

Figure \href{Color online}{\textcolor{blue}{7(a)}} shows diffraction pattern recorded on E4 at 2~K with zero magnetic f\hspace*{.5pt}ield around the (0 0 3/2) reciprocal space position, using projection on the $\omega{-}\upsilon$ plane, where $\omega$ is the rotational angle of the sample and $\upsilon$ is the deviation angle from the scattering plane. As it is evident, the original magnetic Bragg reflection splits into six reflections indexed by three propagation vectors $q_{1}$ = ($q_{h} \hspace*{1.5pt} q_{h} \hspace*{1.5pt} 3/2$), $q_{2}$ = (-$q_{h}\hspace*{1.5pt} 2q_{h} \hspace*{1.5pt} 3/2$), and $q_{3}$ = (-2$q_{h} \hspace*{1.5pt} q_{h} \hspace*{1.5pt} 3/2$) (and associated opposite vectors). Using the UB matrix ref\hspace*{.5pt}ined from very few nuclear Bragg reflections and positions of the maxima, it follows that $q_{h}$ = 0.027 (1) is in agreement with the literature \textcolor{blue}{\cite{Svergun1987,Cameron2016}}. The six magnetic reflections can be indexed as ($-$0.027 $-$0.027 3/2), ($-$0.054 0.027 3/2), ($-$0.027 0.054 3/2), (0.027 0.027 3/2), (0.054 $-$0.027 3/2) and (0.027 $-$0.054 3/2). These propagation vectors are incommensurate with the crystal structure. The projection of Fig.~\href{Color online}{\textcolor{blue}{7(a)}} on the rotational axis, $\omega$ is shown in Fig.~\href{Color online}{\textcolor{blue}{7(b)}}, suggesting that these six reflections have inhomogeneous intensity distributions. 
The existence of six magnetic propagation vectors at low temperatures raises question whether one deals with a homogeneous state where the Ni magnetic moments are modulated in the whole volume of the sample by all the propagation vectors or whether one deals with spatially disjoint domains, each having one propagation vector. However, from the fact that, for instance, the reflection described by the propagation vector $q_{3}$ is by about 50 $\% $ more intense than the $-q_{2}$ reflection (see Fig.~\href{Color online}{\textcolor{blue}{7(a)}}), the former scenario can be ruled out. This conf\hspace*{.5pt}irms that the magnetic structure consists of volume-separated domains. 

From a symmetry point-view, the most distinctive properties of the incommensurate spiral state are from the 120\textdegree\ structures. The ground state displays threefold degeneracy, \textit{i.e.}, there are three equivalent directions of wave vectors on the lattice. A f\hspace*{.5pt}irst-order transition is associated with a breaking of the threefold C$_{3}$ lattice symmetry. The ordered state is a single-\textit{q} state with three equivalent wave vector directions. This threefold degeneracy could also be in principle a source of exotic ordered states, e.g., various types of multiple-\textit{q} state where multiple wave vectors coexist. The helix structure is realized at the magnetic ground state below $T_{{\rm N}}$, where spins rotate within the plane perpendicular to the magnetic modulation vector.

Figure \href{Color online}{\textcolor{blue}{8(a)}} shows the temperature dependence of the diffracted intensities projected on the $\omega$-axis along with projections on the $\omega{-}\upsilon$ plane (Fig.~\href{Color online}{\textcolor{blue}{8(b)}}) recorded at various temperatures. As can be seen, the six incommensurate reflections collapse at $T_{{\rm m}}$ to a single (0 0 3/2) reflection. Figure \href{Color online}{\textcolor{blue}{8(c)}} illustrates the temperature dependence of the intensity around (0 0 3/2). Total diffracted intensity is included, irrespective of whether originating from a commensurate or incommensurate magnetic state. The absence of any anomaly around $T_{{\rm N}}$ suggests that the (0 0 3/2) reflection just splits into six reflections at lower temperature without any signif\hspace*{.5pt}icant change in the magnetic structure/magnetic moment magnitude.

Symmetry analysis for both types of propagation vectors (commensurate and incommensurate) leads to conclusion that Ni magnetic moments are oriented either along the \textit{c}-axis or perpendicular to it. The fact that we do observe intensities at (0 0 3/2) position, which suggests that the moments are not along the trigonal axis and are oriented within the plane. However, due to high symmetry of the crystal structure, it is not possible to determine the direction of these moments within the (\hbox{\hbox{$a$--$b$}}) basal plane.

The best f\hspace*{.5pt}it to data shown in Fig.~\href{Color online}{\textcolor{blue}{5(b)}}, i.e., data taken in zero f\hspace*{.5pt}ields at 25~K, leads to Ni moment magnitude of 2.8 (2) $\mu_{\mathrm{B}}$. This value is somewhat larger with respect to the literature data \textcolor{blue}{\cite{Robbins1588,Meriel1680}}. The feasible magnetic structures are shown in Fig.~\href{Color online}{\textcolor{blue}{9}}. It is noteworthy here that this is one of the several possible magnetic structures that agree equally with the data. The difference between solution shown in Fig.~\href{Color online}{\textcolor{blue}{9(a and b)}} is that Ni moments are collinear in Fig.~\href{Color online}{\textcolor{blue}{9(a)}} and smaller than in Fig.~\href{Color online}{\textcolor{blue}{9(b)}}. The best f\hspace*{.5pt}it to data taken at 2~K, 0 T as shown in Fig.~\href{Color online}{\textcolor{blue}{5(c)}} using six propagation vectors $q_{1}$ = ($-$0.027 $-$0.027 3/2), $q_{2}$ = ($-$0.054 0.027 3/2), $q_{3}$ = ($-$0.027 0.054 3/2), $q_{4}$ = (0.027 0.027 3/2), $q_{5}$ = (0.054 $-$0.027 3/2), and $q_{6}$ = (0.027 $-$0.054 3/2) leads to Ni magnetic moment of 3.0 (3) $\mu_{\mathrm{B}}$. The direction of Ni moments are changing within the basal plane in an incommensurate manner, making with respect to each other about $\alpha$ = 9.5$^\circ$ as one moves within the plane along \textit{a}-axis as illustrated in Fig.~\href{Color online}{\textcolor{blue}{9(c)}}. This solution is in good agreement with the literature data \textcolor{blue}{\cite{Robbins1588,Meriel1680,Terrier1982}}. The population of six magnetic domains is about $18\%$:$10\%$:$22\%$:$13\%$:$19\%$:$18\%$.

In this section, we would elaborate in more detail about the spin-flop mechanism across the incommensurate phase that is induced in the presence of an external magnetic f\hspace*{.5pt}ield (H $\geq$ 4~T) applied within the (\hbox{\hbox{$a$--$b$}}) plane. The high temperature magnetic phase is retained, which suggests that only the magnetic reflections indexable with $q_{c}$ = (0 0 3/2) are present (see Fig.~ \href{Color online}{\textcolor{blue}{5(b) $\&$ 5(d)}}). This is also shown in Fig.~ \href{Color online}{\textcolor{blue}{10(c)}}, which shows the f\hspace*{.5pt}ield dependence of the intensity projected on $\omega$-axis with increasing f\hspace*{.5pt}ield. Merely the intensities of AF reflections are reduced with respect to the zero f\hspace*{.5pt}ield value, pointing to a smaller staggered magnetic moment of 2.5 (2) $\mu_{\mathrm{B}}$. No signif\hspace*{.5pt}icant ferromagnetic component could be resolved. 

In Fig.~\href{Color online}{\textcolor{blue}{10(b)}}, the temperature dependence of the intensity around (0 0 3/2) position is measured with increasing temperature in a f\hspace*{.5pt}ield of 2 T applied within the (\hbox{\hbox{$a$--$b$}}) plane. As can be seen, the lower transition shifts with respect to zero f\hspace*{.5pt}ields toward lower temperatures while the upper transition stays approximately unchanged that in well agreement with magnetization data. As the f\hspace*{.5pt}ield increases at 2~K above the critical f\hspace*{.5pt}ield of about 3 T, only the magnetic reflections indexable with $q_{c}$ = (0 0 3/2) are present. In Fig.~\href{Color online}{\textcolor{blue}{10(c)}}, the temperature dependence of the intensity around (0 0 3/2) position measured with increasing temperature in a f\hspace*{.5pt}ield of 5 T applied within the (\hbox{\hbox{$a$--$b$}}) plane is shown. Also, in this case, the $T_{{\rm N}}$ remains equal to approximately 44~K. This transition can be interpreted as in the presence of weak anisotropic energy, at a critical magnetic f\hspace*{.5pt}ield the two sub-lattice magnetization rotates suddenly to a direction perpendicular to the easy magnetization direction, i.e., (\hbox{\hbox{$a$--$b$}}) basal plane.

\subsection{\label{sec:phase}Phase diagram}

The magnetic phase diagram of NiBr$_{2}$ near critical temperature $T_{{\rm m}} (T_{{\rm N}})$ has been drawn from the measurements of bulk magnetization, susceptibility, and neutron diffraction. A decrease of the magnetic phase boundary at $T_{{\rm N}}$ results in a substantial increase of the satellite \textit{q}-vector. As shown in the Fig.~\href{Color online}{\textcolor{blue}{11}}, for an intermediate temperature range in the (\hbox{$a$--$b$}) plane, a commensurate magnetic structure between 45~K and 22~K was found by a least-squares ref\hspace*{.5pt}inement to be modulated with moments lying in the (\hbox{$a$--$b$}) plane. At lower temperatures and moderate magnetic f\hspace*{.5pt}ield in the (\hbox{$a$--$b$}) plane, an incommensurate magnetic structure between 22~K and 4~K was found by a least-squares ref\hspace*{.5pt}inement to be modulated with moments lying in the (\hbox{$a$--$b$}) plane with angle of $\gamma$ = 9.6\textdegree\ . The magnetic phase diagram for an external f\hspace*{.5pt}ield applied in the \textit{c}-axis direction was also determined that conf\hspace*{.5pt}irms that upon reducing temperature, one commensurate magnetic structure phase can evolve. The inset shows the extended f\hspace*{.5pt}ield range that suggests for the f\hspace*{.5pt}ield applied within the (\hbox{$a$--$b$}) plane, yet another f\hspace*{.5pt}ield-induced transition around 60 T.

This behavior in $\chi(H, T)$ may be a def\hspace*{.5pt}ining characteristic of such transitions, mainly reflecting the unusual combination of ferromagnetic and antiferromagnetic interactions in the NiBr$_{2}$ system that leads to the complicated influence of applied f\hspace*{.5pt}ields on the changing spin conf\hspace*{.5pt}igurations along the commensurate--incommensurate boundary. Indeed, it is not a universal characteristic of systems with competing ferromagnetic and antiferromagnetic interactions, even when these lead to metamagnetic transitions. The present results thus suggest further study to determine whether such a response is a unique along spin reorientation boundaries-or their analogue occurring in other systems. As Rastelli {et al.} \textcolor{blue}{\cite{Reatto979}} predicted, when $J_{nn}$ and $J_{nnn}$ are both positive, the lattice is ferromagnetic, but when either or both are negative, helical or antiferromagnetic. The helical magnetic structure found in NiBr$_{2}$ at low temperature results from a very delicate and fortuitous balance of the various $J_{i}$. Elevated temperature forces the system across boundary from incommensurate to commensurate phase. The interplay between competing interactions lead to amplitude-modulated magnetic structures with a periodicities that do not match the periodicity of crystal lattice \textcolor{blue}{\cite{Shirane2070,Noda2105,Vekua2206,Randy}}.

Here, it is noteworthy to mention that one of our motivations to undertake the study was to identify whether NiBr$_{2}$ is hosting skyrmions. We have used the SANS method with field applied within the basal plane, similar to studies performed on MnSi \textcolor{blue}{\cite{Muhlbauer}}. We have surveyed a large range of fields up to 4 T and temperatures, focusing mainly around magnetic phase transition lines as shown in Fig.~\href{Color online}{\textcolor{blue}{11}}. Within the given coverage range we did not find any signatures of skyrmions.

\subsection{\label{sec:summmary}SUMMARY AND CONCLUSIONS}
In conclusion, by measuring the magnetic properties and neutron diffraction of model spin triangular system of NiBr$_{2}$, we demonstrated that the variation of uniform exchange, saturated magnetic moment, and helical wave vector systematically as a function of the temperature and applied magnetic f\hspace*{.5pt}ield. The changes we measure in the magnetic properties are dominated by the variation of anisotropy and exchange interactions. These observations are supported by the models as predicted by Rastelli {et al.} \textcolor{blue}{\cite{Reatto979}}, after simplif\hspace*{.5pt}ied assumption of mutual competition in neighboring exchange interactions. Despite this, we demonstrated a magnetic f\hspace*{.5pt}ield-induced phase transition across the incommensurate phase, which entirely transform the low-temperature incommensurate phase into the high temperature commensurate spin structure. This behavior of incommensurate phase is purely governed by spin-flop transition. On the methodological side, our work demonstrates that combining state-of-the-art neutron scattering experiments with magnetization measurements, we were enable to extract def\hspace*{.5pt}initive microscopic information from the spin triangular system of NiBr$_{2}$. These f\hspace*{.5pt}indings accelerate the search for exotic quantum states in helimagnetic systems through the screening of many related materials having direct technological implications. As a conclusion, one may foresee that the accurate understanding the role of neighboring exchange interactions in the triangular spin systems paves the way to exploratory research on f\hspace*{.5pt}ield-induced phase transitions where the choice of spin helix and topological magnetic texture will be a tool to improve the performances.

\begin{acknowledgments}
We would like to thank M. Reehuis, F. Yokaichiya and U. Keiderling for technical help with the neutron data collection. Authors would like to acknowledge Dr. Abhisekh Singh, CIFC, IIT (BHU) for assisting the magnetic measurments. SKM acknowledges the funding from the IIT(BHU) with grant no. IIT(BHU)/R\&D/SM/2016-17/4453. 
\end{acknowledgments}

\clearpage
Figure Captions

\begin{figure}[h!]
	\textcolor{black}{\caption{(Color online) (a) Schematic representation of the crystal structure of NiBr$_{2}$. Green and accent small spheres represent Ni and Br ions in a layered triangular lattice, respectively. $a$, $b$ and $c$ are the principal axes of the hexagonal cell where $c$ is the high-symmetry axis. (b) and (c) depict the collinear antiferromagnetic (AF) $(T_{{\rm m}} <T <T_{{\rm N}})$ and helical spin $(T> T_{{\rm m}})$ structures, respectively. Solid arrows indicate spin directions on Ni$^{2+}$ ions. For simplicity, bromide ions are not shown.}}
	\label{Figure:1}
\end{figure}

\begin{figure}[h!]
	\textcolor{black}{\caption{(Color online) Temperature dependence of the magnetic susceptibility of NiBr$_{2}$ measured with f\hspace*{.5pt}ield was applied parallel to (a) (\hbox{\hbox{$a$--$b$}}) basal plane and (b) along the \textit{c}-axis, respectively. The insets show the temperature dependence of the inverse magnetic susceptibility $1/\chi(T)$ measured at 14 T in both respective panels.The red solid line is the best f\hspace*{.5pt}it of $1/\chi(T)$ to a modif\hspace*{.5pt}ied Curie--Weiss law.}}
	\label{Figure:2}
\end{figure}

\begin{figure}[h!]
	\textcolor{black}{\caption{(Color online)(a) Field dependence of the NiBr$_{2}$ magnetization applied perpendicular to the \textit{c}-axis for various temperatures. The full f\hspace*{.5pt}ield range ($\pm$ 14 T) of M--H curves are shown in the inset. (b) Field dependence of the NiBr$_{2}$ magnetization applied along the \textit{c}-axis measured at various representative temperatures between 3~K and 60~K.}}
	\label{Figure:3} 
\end{figure}

\begin{figure}[h!]
	\textcolor{black}{\caption{(Color online) (a) The temperature dependence of the ac susceptibility, $\chi$(H,T) of NiBr$_{2}$ in the vicinity of the lower transition(s), near 50~K, measured in various static applied f\hspace*{.5pt}ields between 0.1 and 2.0 T. Isothermal ac susceptibility $\chi$(H,T) measured over the range $5~K\leq T\leq80~K$, $0.0\leq H\leq3.0 T$. (b) The hysteresis regime Fig.~\href{Color online}{\textcolor{blue}{4(b)}} corresponds to the incommensurate phase, $5~K\leq T\leq 25~K$. For higher temperature $25~K\leq T\leq 80K$, $\chi$(H,T) shows a straight line conf\hspace*{.5pt}irming the compensated antiferromagnetic phase area. (c) Comparison of ac (blue solid line) and dc (red solid line) magnetic susceptibilities across the transitions. Temperature derivatives $\dv{\chi'_{ac}}{T}$ and $\dv{\chi_{dc}}{T}$ are depicted in the inset.}}
	\label{Figure:4}
\end{figure}

\begin{figure}[h!]
	\textcolor{black}{\caption{(Color online) Diffraction patterns of the NiBr$_{2}$ single crystal in the (\textit{hhl}) plane at mentioned conditions: (a) 50~K, 0 T, (b) 26~K, 0 T, (c) 2~K, 0 T and (d) 2~K, 4 T applied within the (\hbox{\hbox{$a$--$b$}}) basal plan. Insets show the highlighted area around the (0 0 3/2) reciprocal space.}}
	\label{Figure:5}
\end{figure}

\begin{figure}[h!]
	\textcolor{black}{\caption{(Color online) (a) Rocking curve of the nuclear (003) reflection recorded at 55~K on E4. (b) Rocking curve of the magnetic (0 0 3/2) reflection recorded at 23~K. (c) The temperature dependence of the magnetic (0 0 3/2) Bragg reflection measured upon cooling at the top of the reflection. The solid line through the points just below the $T_{{\rm N}}$ is the best f\hspace*{.5pt}it to the Eq. (4) given in the text. The second-order transition at $T_{{\rm m}}$ takes place at 21.6 (5) K. Just above the $T_{{\rm N}}$, a magnetic diffraction signal due to short-range correlations is visible, which is highlighted with a dashed line.}}
	\label{Figure:6}
\end{figure}

\begin{figure}[h!]
	\textcolor{black}{\caption{(Color online) (a) The diffraction pattern recorded with zero f\hspace*{.5pt}ield around the (0 0 3/2) magnetic reflections in reciprocal space using $\omega{-}\upsilon$ projection. (b) The conventional projection of diffraction pattern on the rotational $\omega$-axis.}}
	\label{Figure:7}
\end{figure}

\begin{figure}[h!]
	\textcolor{black}{\caption{(Color online) (a and b) The temperature dependence of the diffracted signal around the magnetic (0 0 3/2) Bragg reflection projected on the $\omega$-axis along with detection as seen on the 2D detector on E4 at various temperatures in zero f\hspace*{.5pt}ield. (c) Temperature dependence of the intensity around the (0 0 3/2) position integrated in such a way that all six magnetic reflections shown in (b) are included.}}
	\label{Figure:8}
\end{figure}

\begin{figure}[h!]
	\textcolor{black}{\caption{(Color online) Schematic representation of AF structure of NiBr$_{2}$ at 26~K, 0T (a) and 2~K , 0 T (b). (c) A possible solution that agrees with data taken at 26~K, 0T equally well. Only Ni atoms are shown. In all cases, $2\times2\times1$ crystallographic unit cells are shown. The moments in underlying cell along the \textit{c}-axis are reverted.}}
	\label{Figure:9}
\end{figure}

\begin{figure}[h!]
	\textcolor{black}{\caption{(Color online) (a) Field dependence of the intensity around the (0 0 3/2) in $\omega-\upsilon$ reciprocal plane and its projection on $\omega$-axis measured at 2~K with increasing f\hspace*{.5pt}ield. (b) Temperature dependence of the intensity around the (0 0 3/2) measured with f\hspace*{.5pt}ield of 2 T applied within the (\hbox{\hbox{$a$--$b$}}) plane. (c) Temperature dependence of the intensity around the (0 0 3/2) measured with f\hspace*{.5pt}ield of 5 T applied within the (\hbox{\hbox{$a$--$b$}}) plane.}}
	\label{Figure:10}
\end{figure}

\begin{figure}[h!]
		\textcolor{black}{\caption{(Color online) Magnetic phase diagram of NiBr$_{2}$ constructed from magnetic bulk measurements (circles and squares) and neutron diffraction data (stars). In the inset, a crude estimate of the critical f\hspace*{.5pt}ield associated with the $T_{{\rm N}}$ is represented. The dotted line is a guide for an eye.}}
		\label{Figure:11}
\end{figure}

\clearpage

Figure 1
\begin{figure}[h!]
	
	\includegraphics[height=12 cm, width=17 cm]{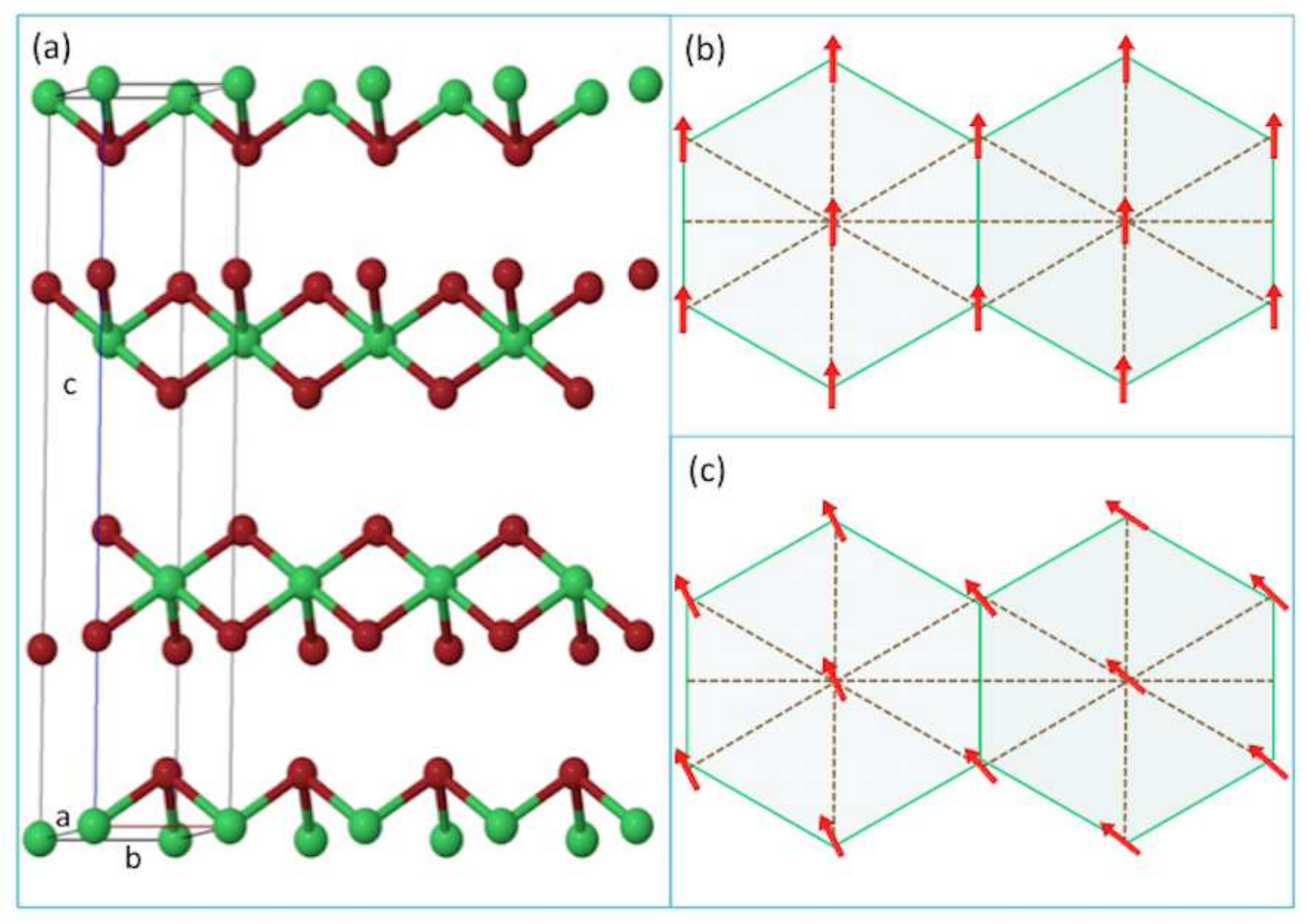}
	\label{Figure:1}
\end{figure}

\clearpage
Figure 2

\begin{figure}[h!]
	\includegraphics[height=22 cm, width=15 cm]{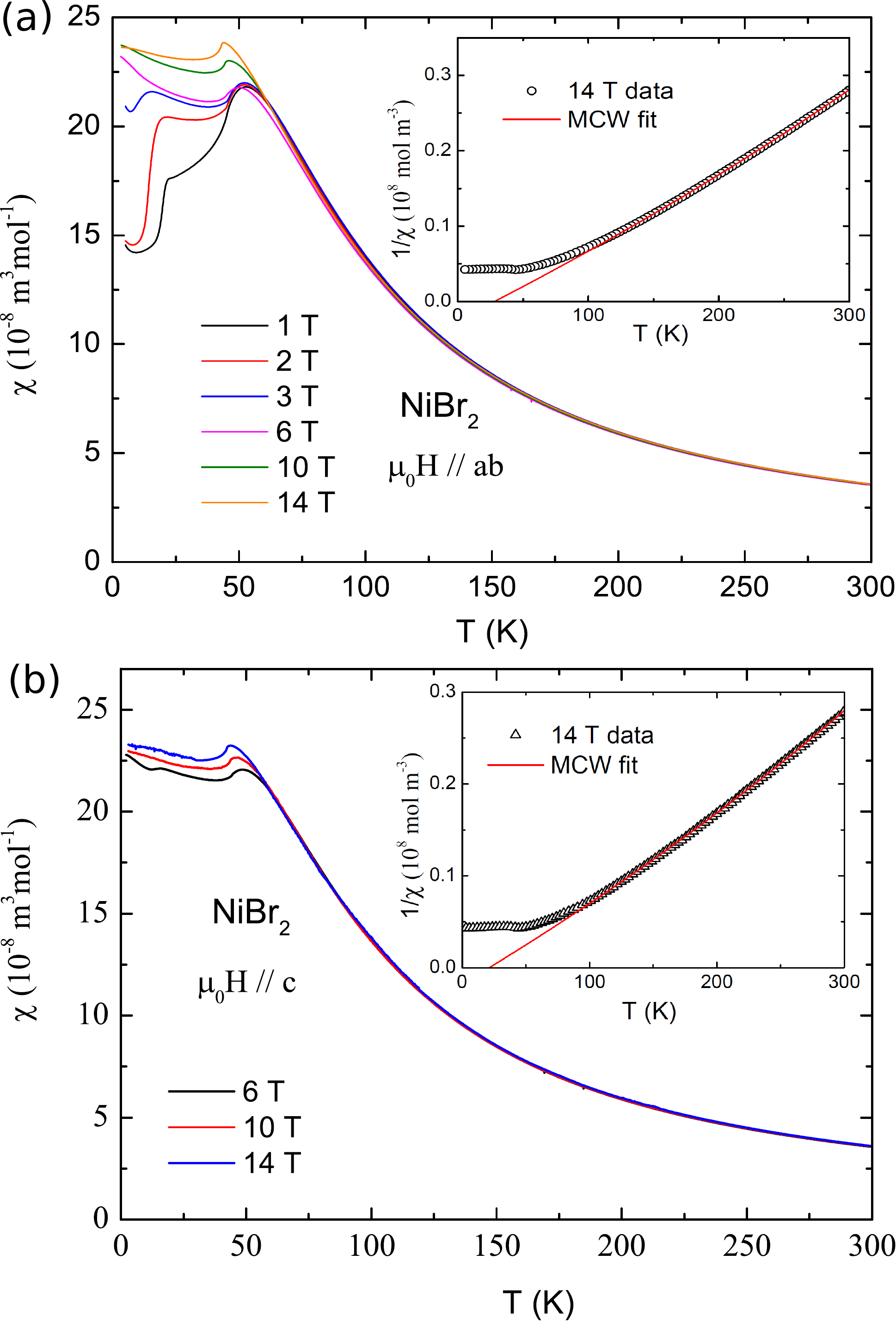}
	\label{Figure:2}
\end{figure}

\clearpage
Figure 3

\begin{figure}[h!]
	\includegraphics[height=20 cm, width=15 cm]{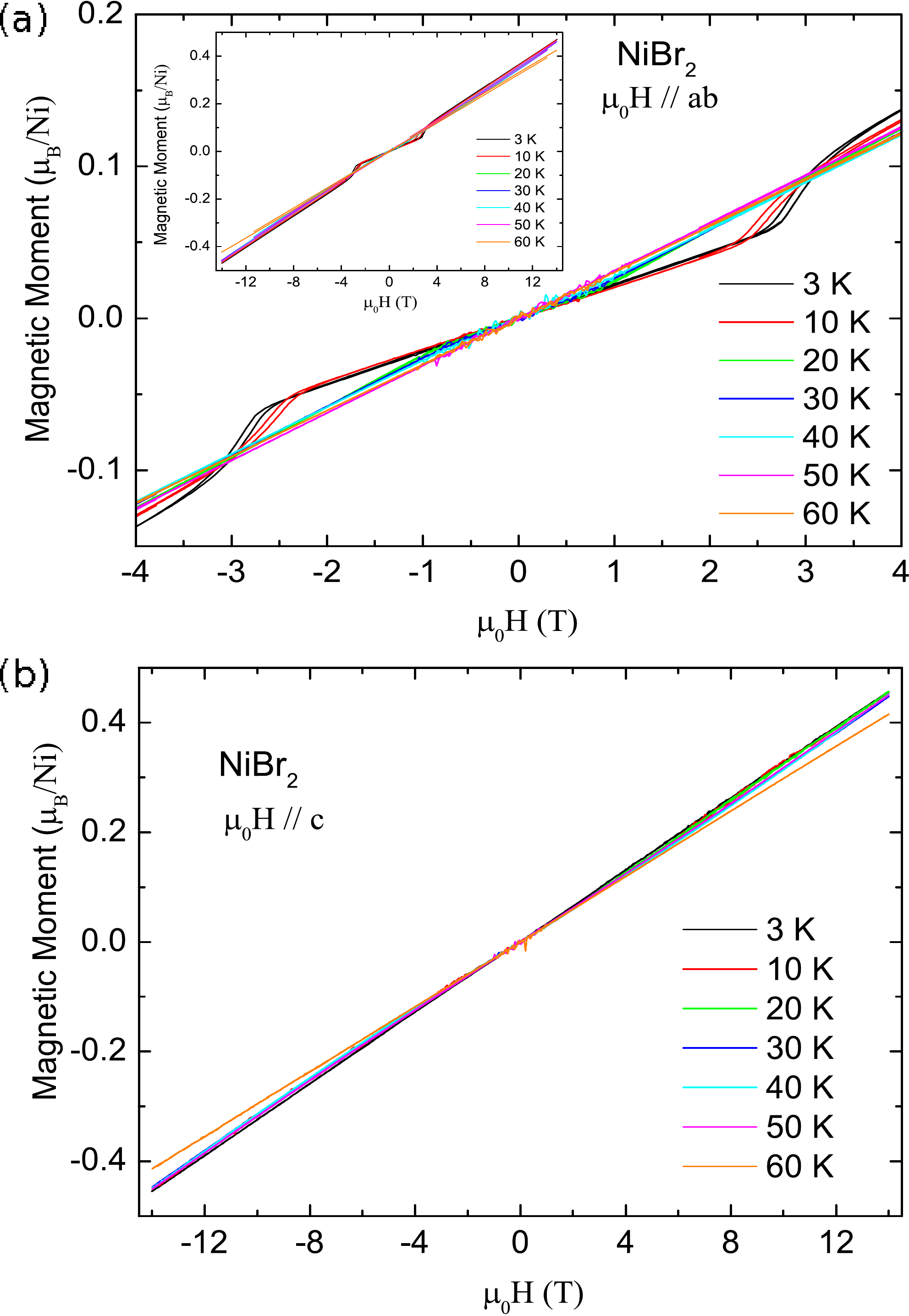}
	\label{Figure:3}
\end{figure}

\clearpage
Figure 4

\begin{figure}[h!]
	\includegraphics[height=22.0 cm, width=12.5 cm]{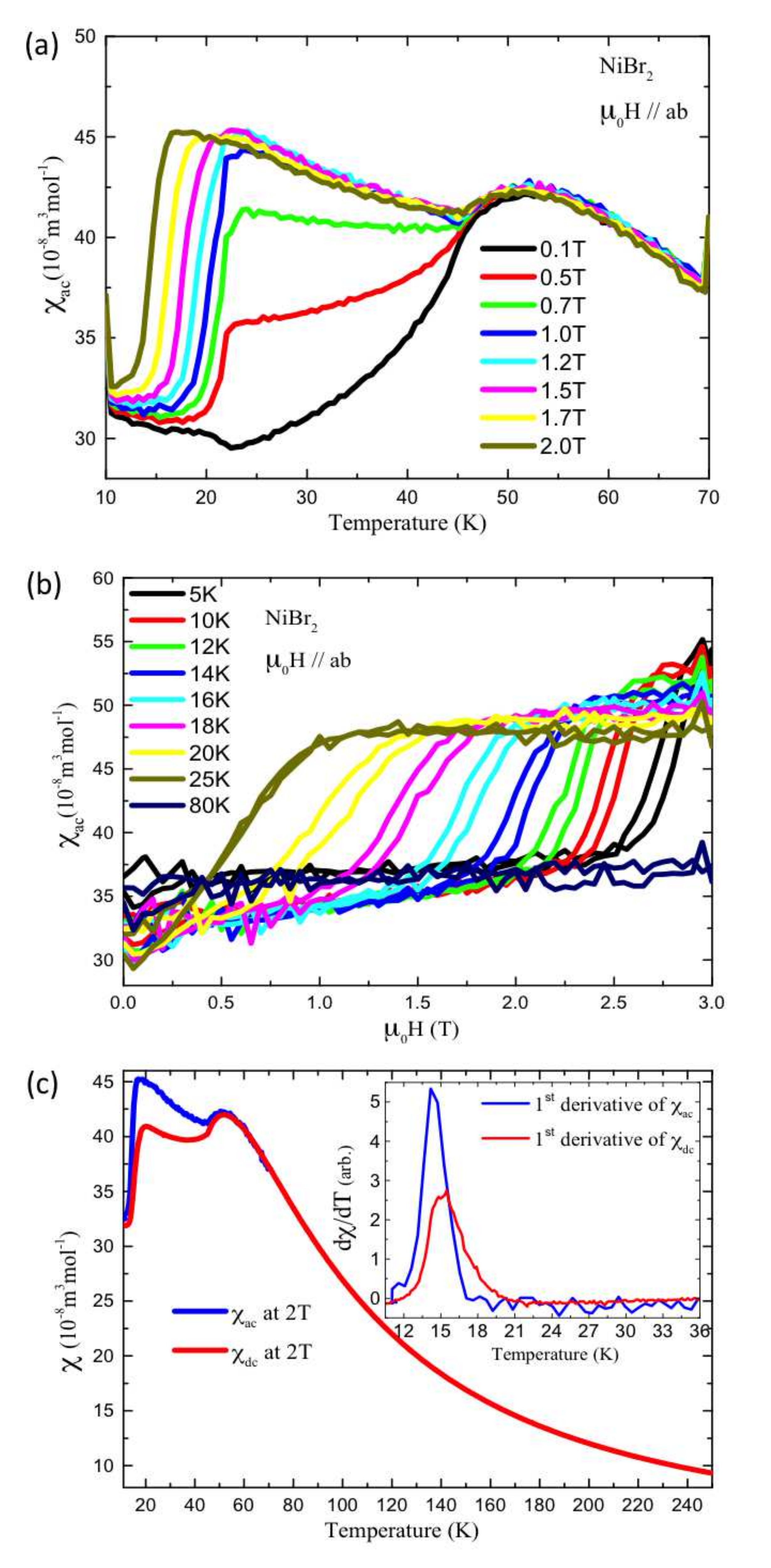}
	\label{Figure:4}
\end{figure}

\clearpage

Figure 5
\begin{figure}[h!]
	\includegraphics[height=15 cm, width=17.0 cm]{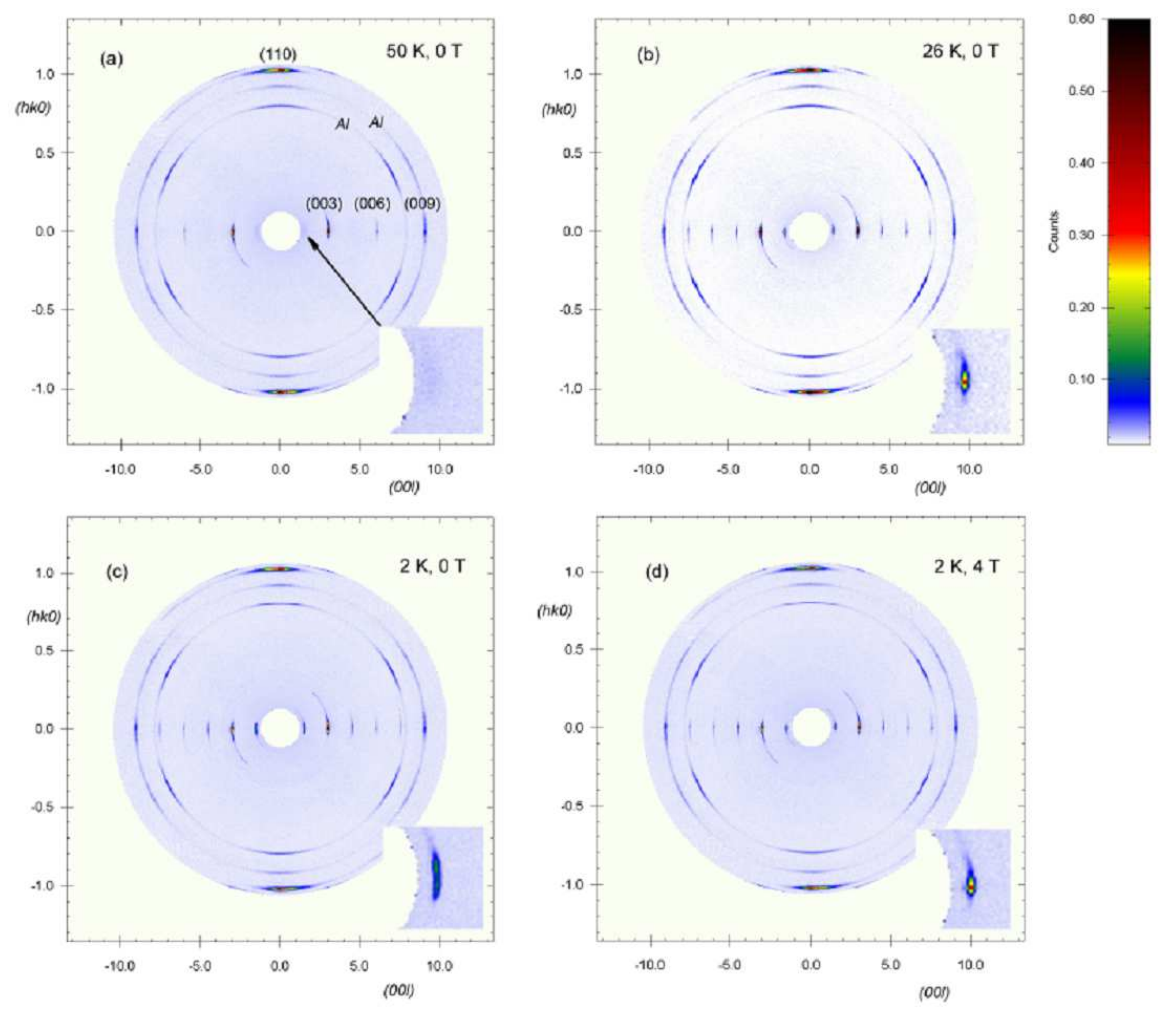}
	\label{Figure:5}
\end{figure}

\clearpage
Figure 6

\begin{figure}[h!]
	\includegraphics[height=7.0 cm, width=11.0 cm]{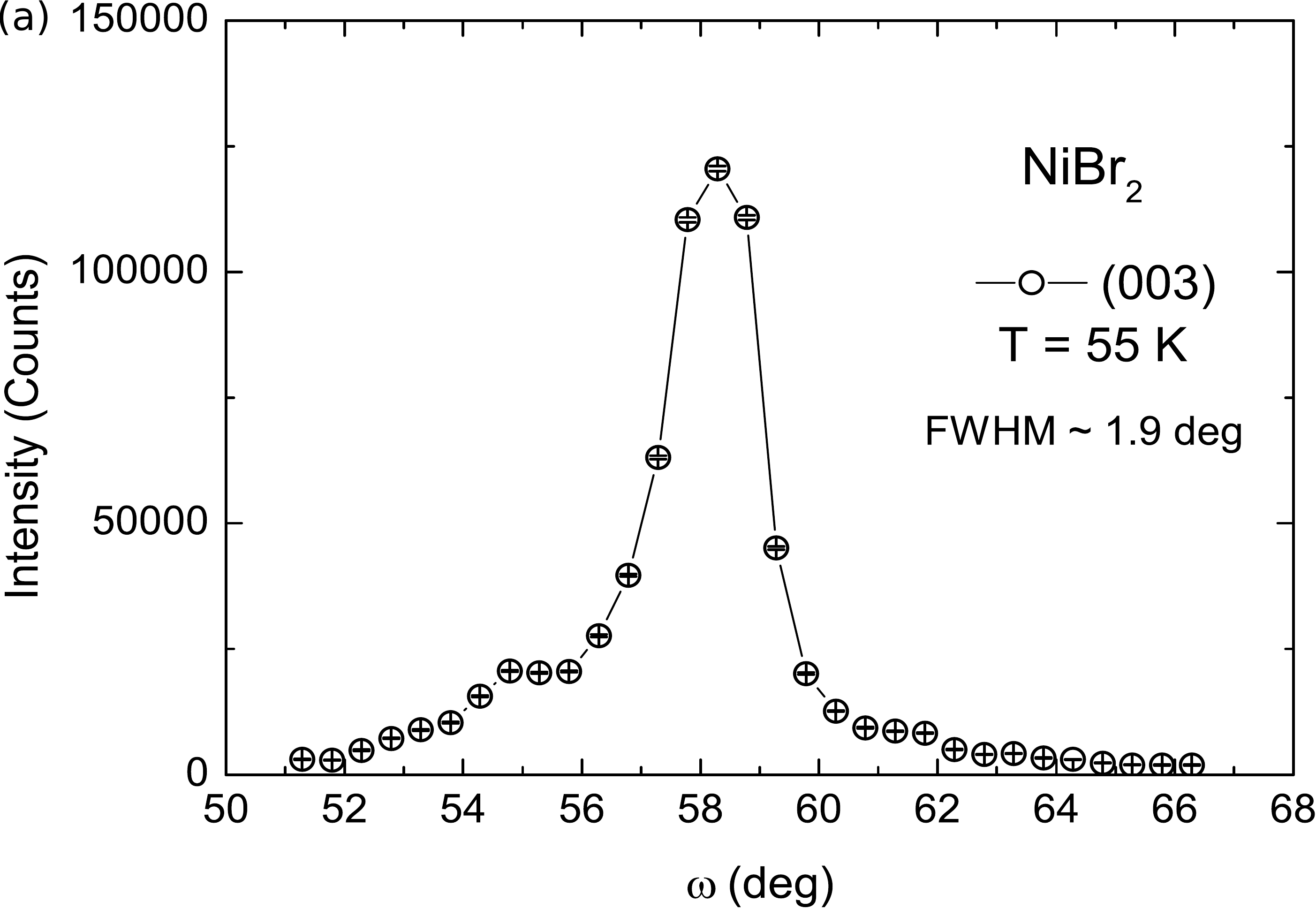}
	\includegraphics[height=7.0 cm, width=10.75 cm]{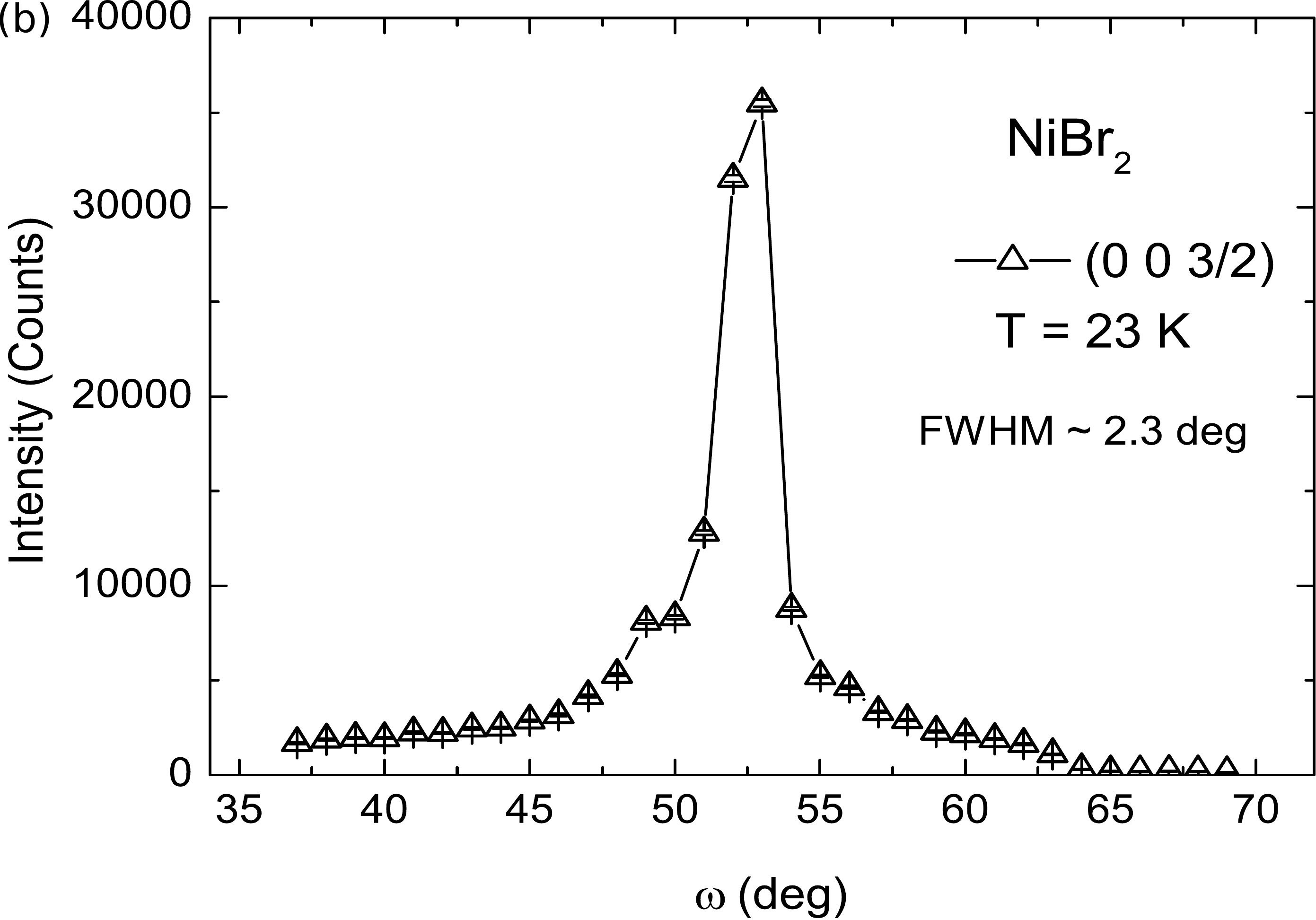}
	\includegraphics[height=7.0 cm, width=11.0 cm]{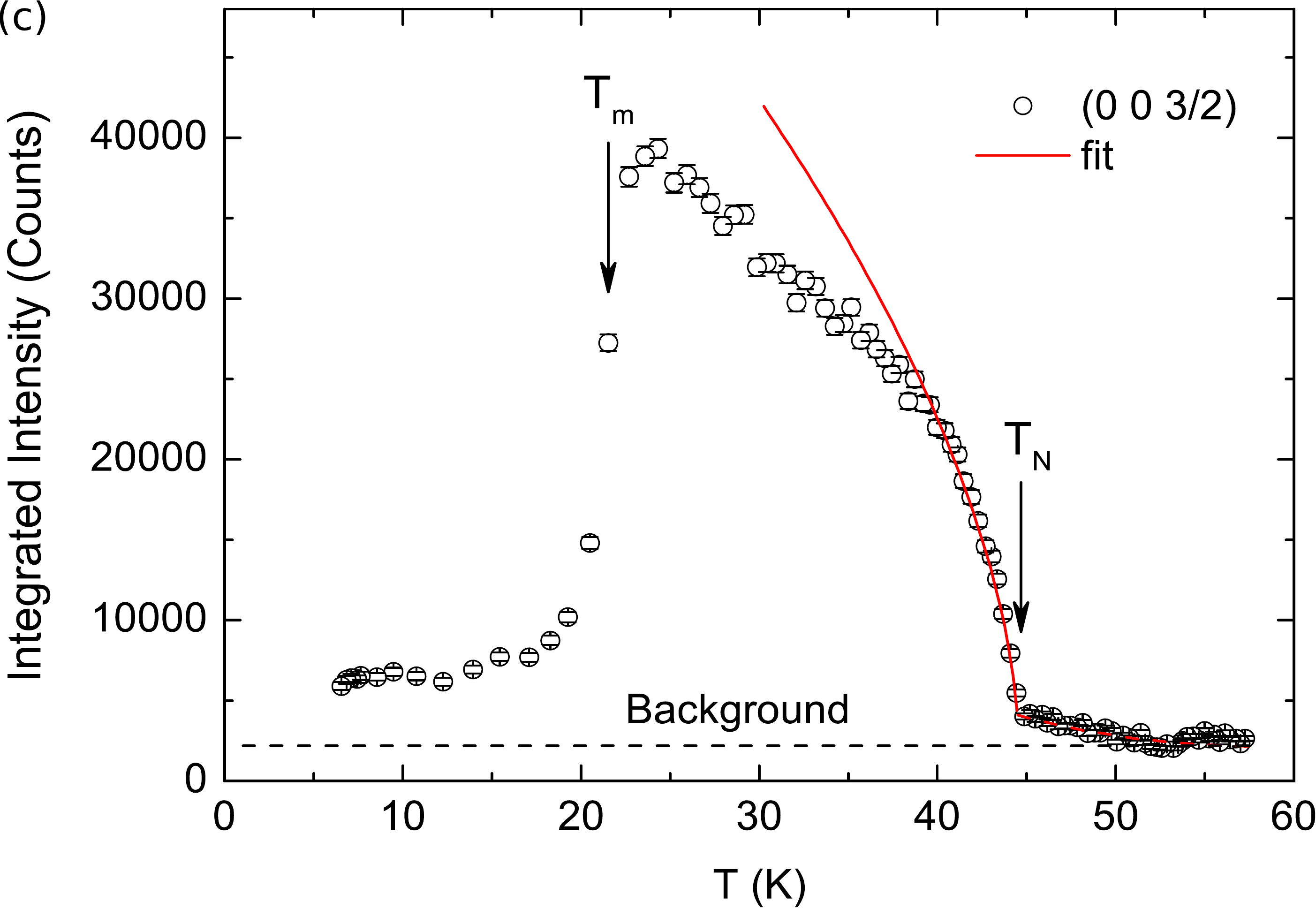}
	\label{Figure:6}
\end{figure}

\clearpage
Figure 7

\begin{figure}[h!]
    \includegraphics[height=22.0 cm, width=16.0cm]{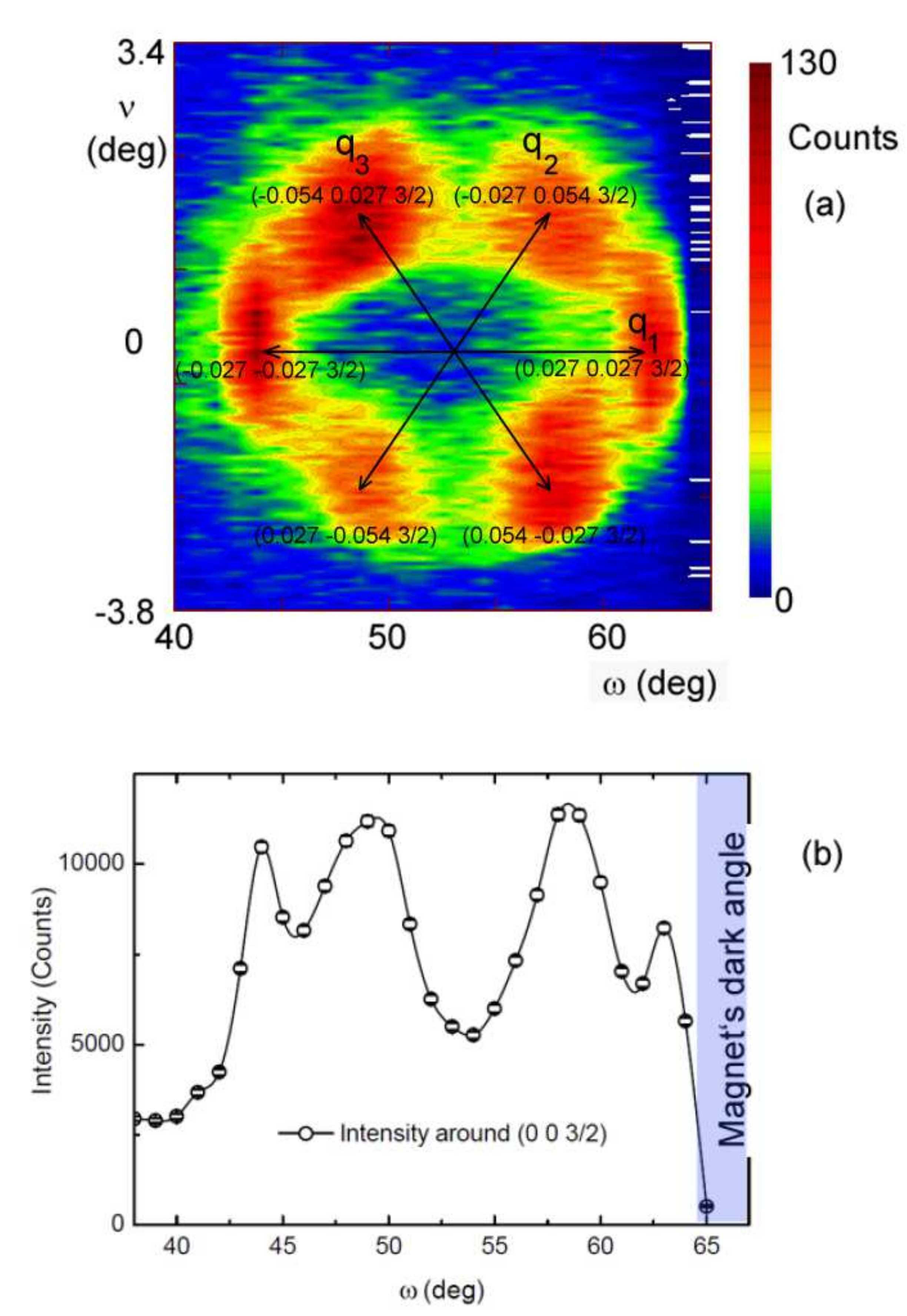}
    \label{Figure:7}
\end{figure}

\clearpage
Figure 8

\begin{figure}[h!]
	\includegraphics[height=12.0 cm, width=14.0 cm]{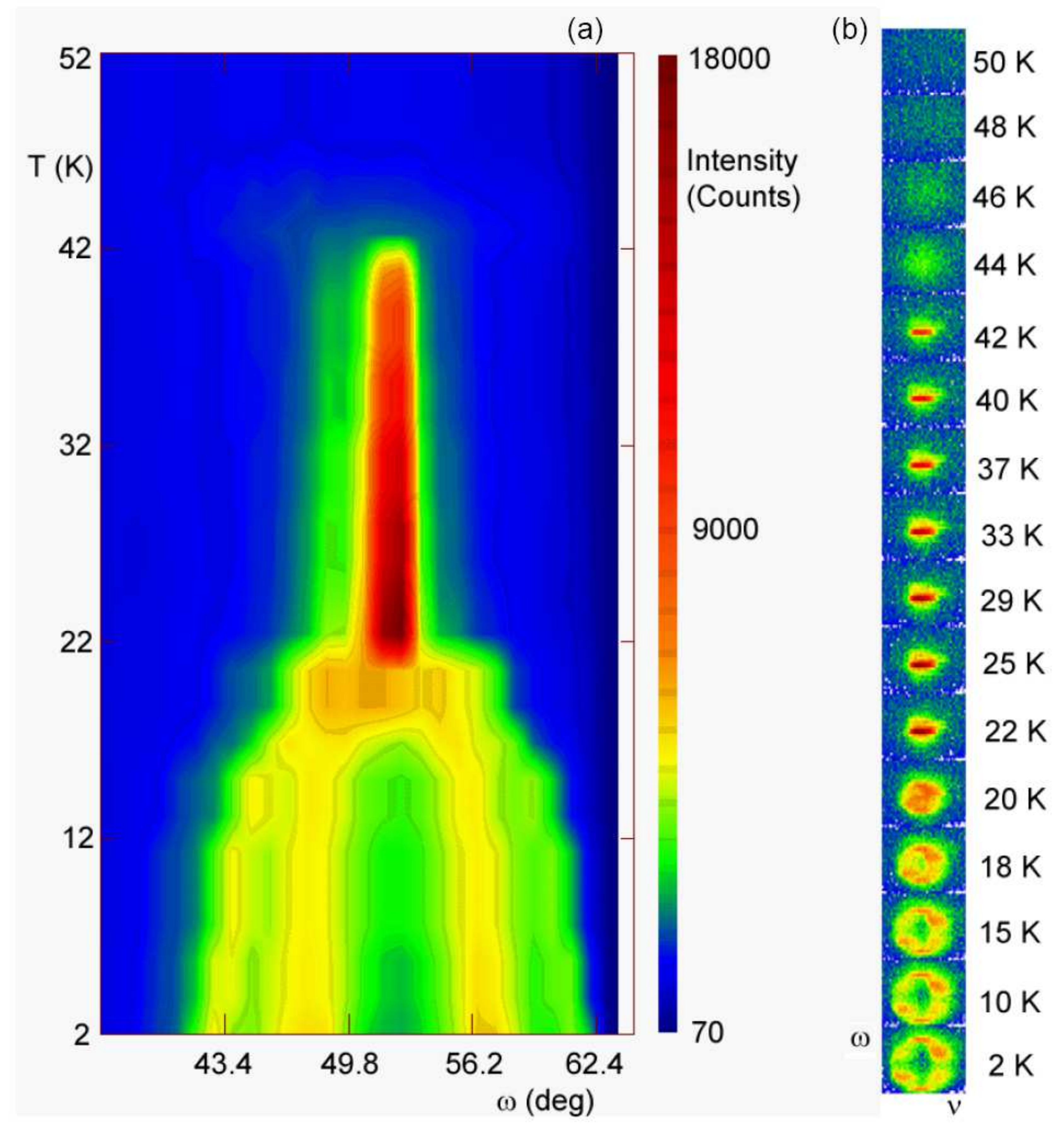}
	 \includegraphics[height=10.0 cm, width=14.0 cm]{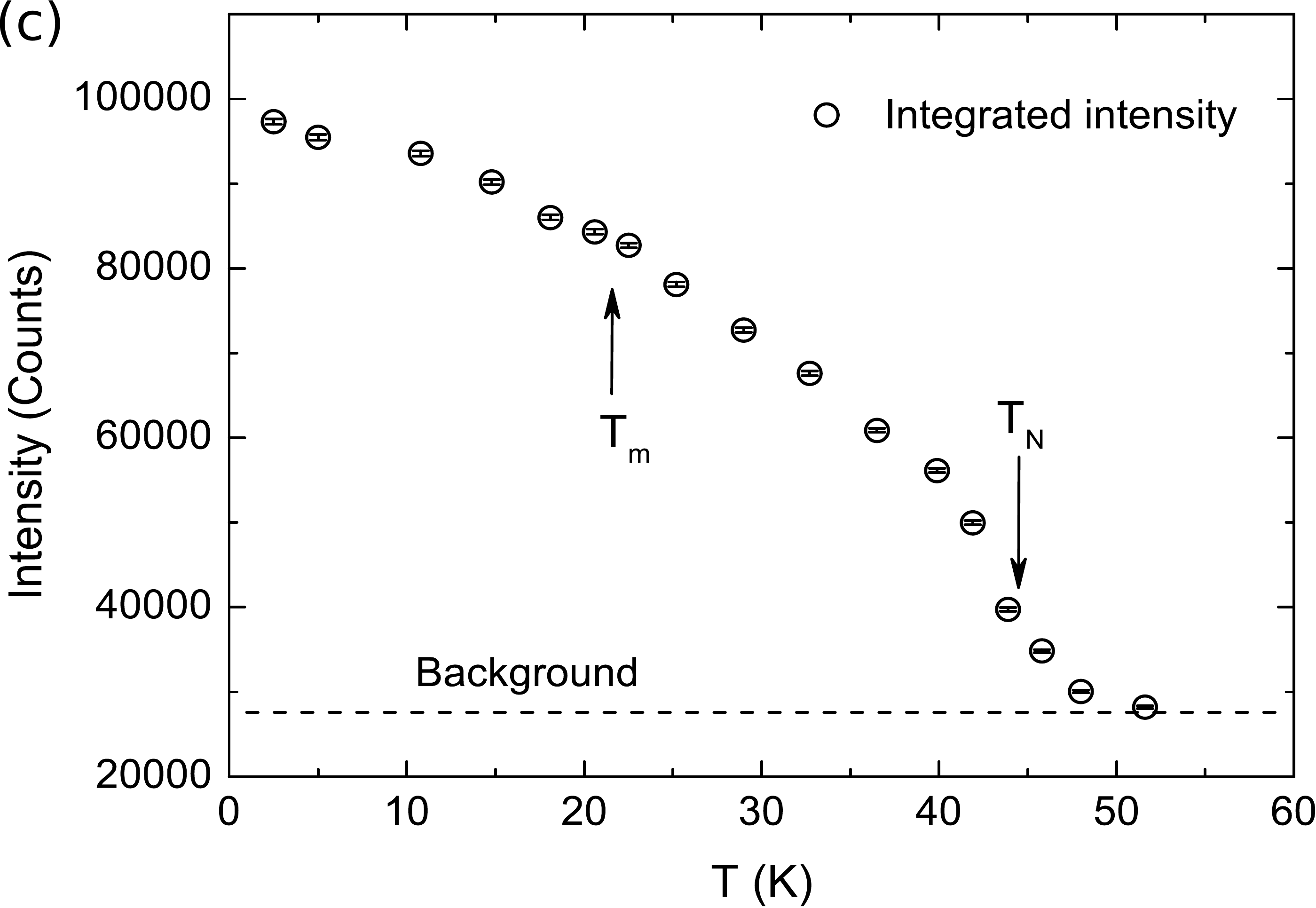}
	\label{Figure:8}
\end{figure}

\clearpage
Figure 9

\begin{figure}[h!]
	 \includegraphics[height=9.5 cm, width=17.5 cm]{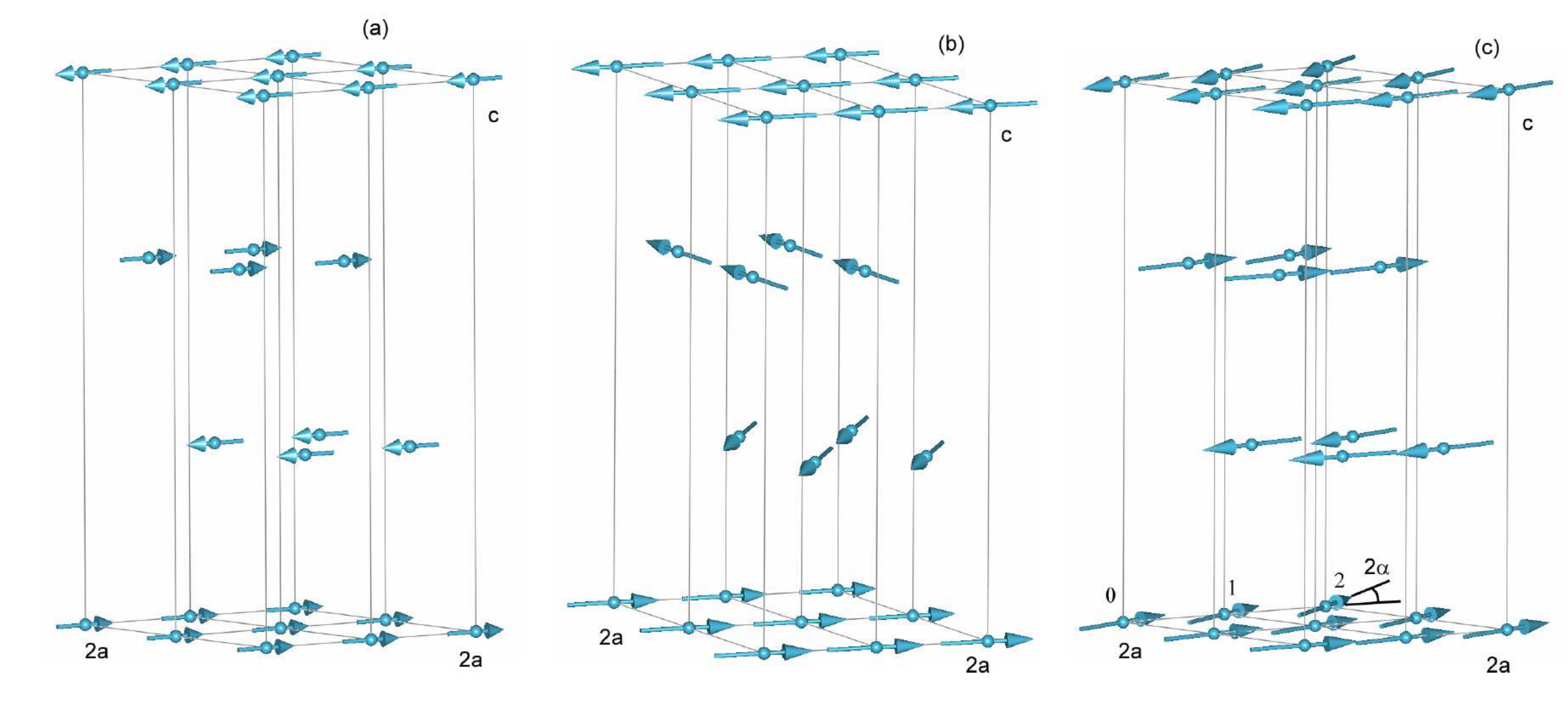}
	\label{Figure:9}
\end{figure}

\clearpage
Figure 10

\begin{figure}[h!]
	\includegraphics[height=10.0 cm, width=8.0 cm]{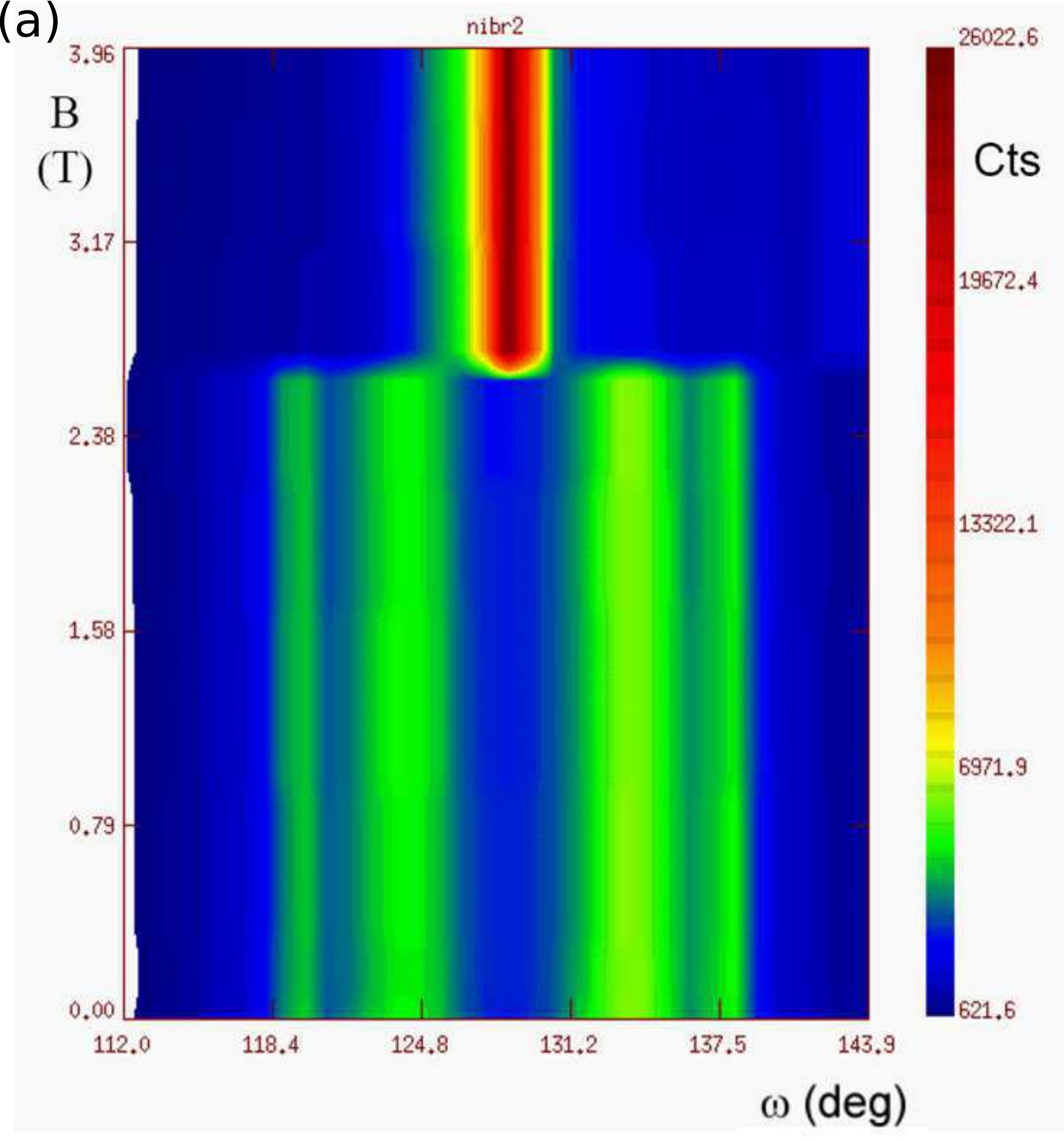}
	\includegraphics[height=10.0 cm, width=8.0 cm]{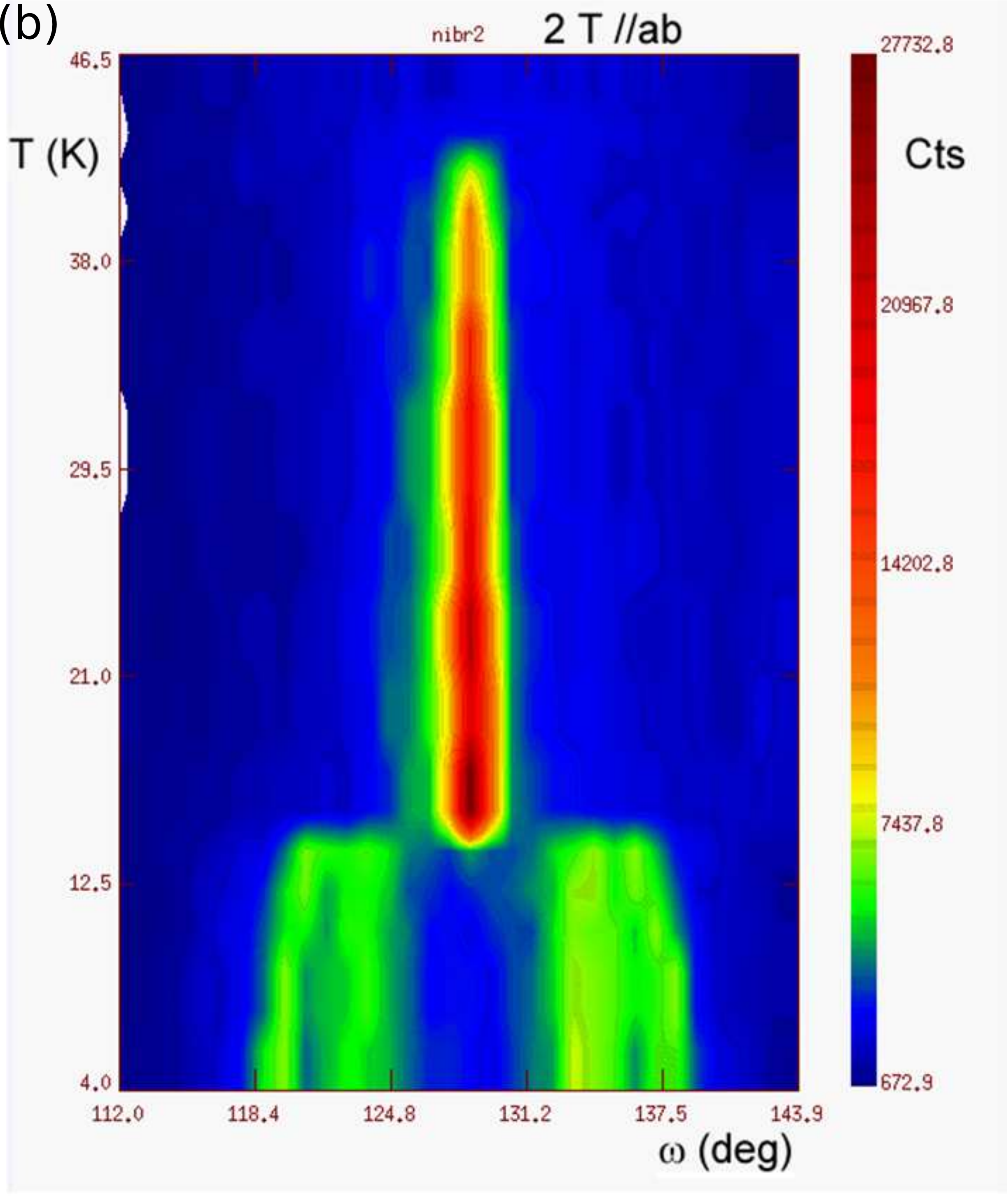}
	\includegraphics[height=10.0 cm, width=8.0 cm]{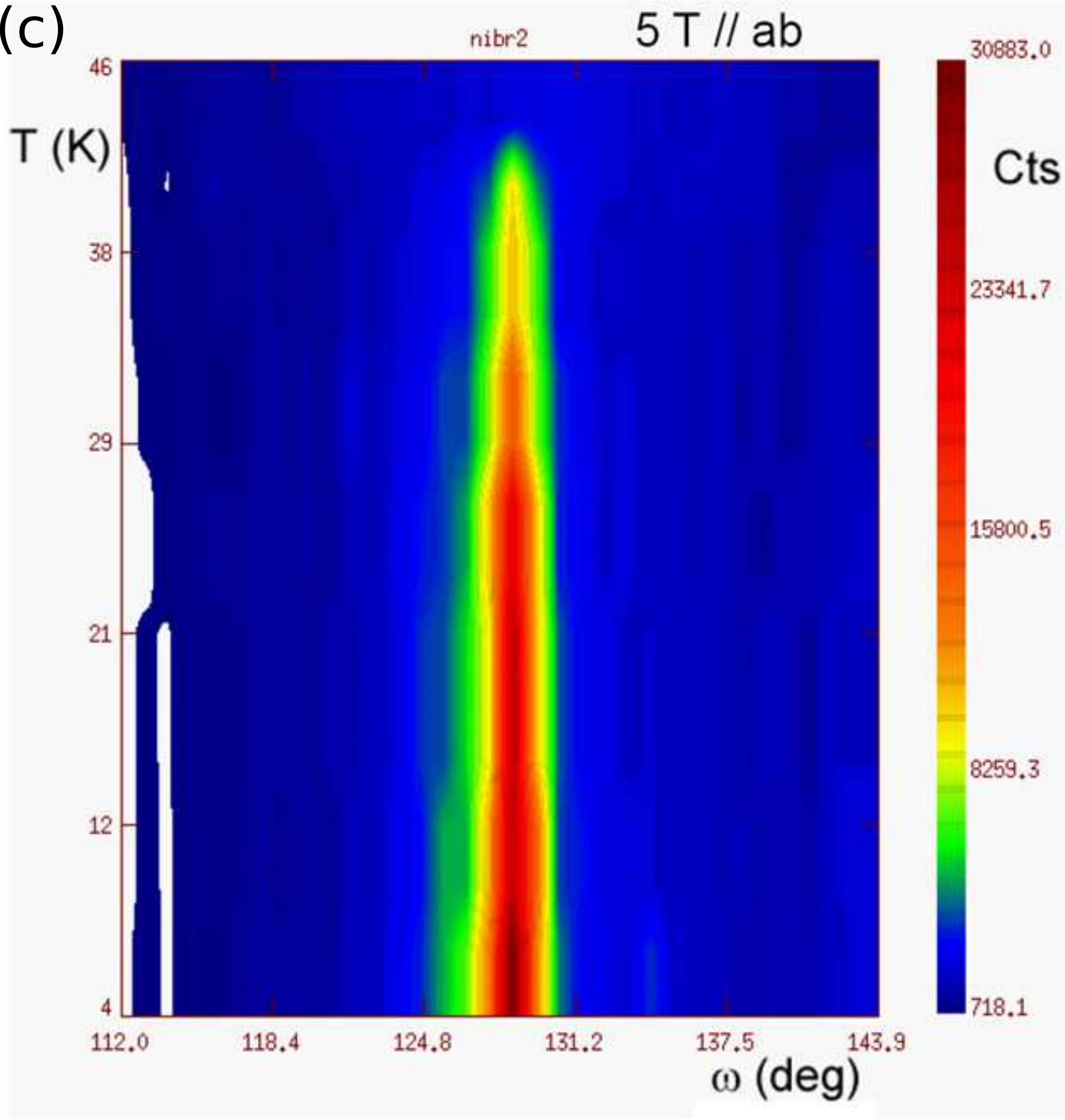}
	\label{Figure:10}
\end{figure}

\clearpage
Figure 11

\begin{figure}[h!]
	 \includegraphics[height=13.0 cm, width=19.5 cm]{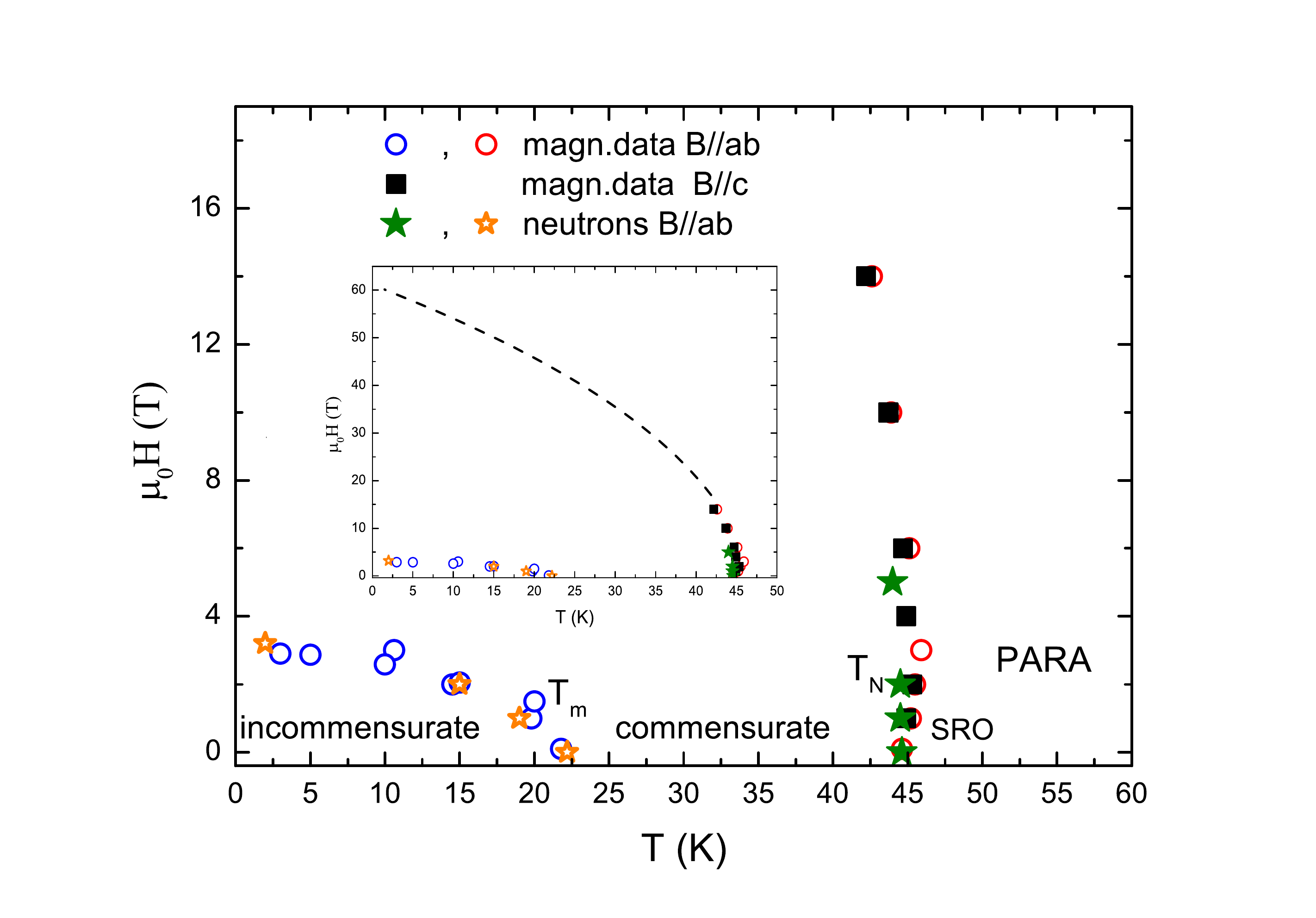}
	\label{Figure:11}
\end{figure} 

\end{document}